\documentclass[aps,prx,article,twocolumn,preprintnumbers,amsmath,amssymb,superscriptaddress, longbibliography]{revtex4}

\date{\today}
\usepackage{epsfig}
\usepackage{cases}
\usepackage{subfigure}
\usepackage{amsmath}
\usepackage{graphicx}
\usepackage{dcolumn}
\usepackage{bm}
\usepackage[colorlinks,linkcolor=blue,hyperindex,CJKbookmarks]{hyperref}
\usepackage{float}
\usepackage{mathtools}
\usepackage{hyperref}
\usepackage{comment}
\usepackage{bbm}
\begin{document}
\newcommand{\Ham}{\mathcal{H}}
\newcommand{\kbf}{\mathbf{k}}
\newcommand{\qbf}{\mathbf{q}}
\newcommand{\Qbf}      {\textbf{Q}}
\newcommand{\lbf}      {\textbf{l}}
\newcommand{\ibf}      {\textbf{i}}
\newcommand{\jbf}      {\textbf{j}}
\newcommand{\rbf}      {\textbf{r}}
\newcommand{\Rbf}      {\textbf{R}}
\newcommand{\Schrdg} {{Schr\"{o}dinger}}
\newcommand{\aband} {{(\alpha)}}
\newcommand{\bband} {{(\beta)}}

\newcommand{\eps} {{\bm{\varepsilon}}}
\newcommand{\probA}      {{\mathsf{A}}}
\newcommand{\HamPump}      {{\Ham_{\rm pump}}}
\newcommand{\HamPr}      {{\Ham_{\rm probe}}}
\newcommand{\timeMax} {{t_{\rm m}}}
\newcommand{\qin} {{\qbf_{\rm i}}}
\newcommand{\qout} {{\qbf_{\rm s}}}
\newcommand{\epsin} {{\eps_{\rm i}}}
\newcommand{\epsout} {{\eps_{\rm s}}}
\newcommand{\win} {{\omega_{\rm in}}}
\newcommand{\wout} {{\omega_{\rm s}}}

\title{Time-Resolved Resonant Inelastic X-Ray Scattering in a Pumped Mott Insulator}
\author{Yao Wang }
 \affiliation{Department of Physics, Harvard University, Cambridge, Massachusetts 02138, USA}
 \author{Yuan Chen }
 \affiliation{Department of Applied Physics, Stanford University, California 94305, USA}
 \affiliation{Stanford Institute for Materials and Energy Sciences, SLAC National Accelerator Laboratory and Stanford University, Menlo Park, California 94025, USA}
\author{Chunjing Jia}
\affiliation{Stanford Institute for Materials and Energy Sciences, SLAC National Accelerator Laboratory and Stanford University, Menlo Park, California 94025, USA}
\author{Brian Moritz}
\affiliation{Stanford Institute for Materials and Energy Sciences, SLAC National Accelerator Laboratory and Stanford University, Menlo Park, California 94025, USA}
\affiliation{Department of Physics and Astrophysics, University of North Dakota, Grand Forks, North Dakota 58202, USA}
\date{\today}
\author{Thomas P. Devereaux}
\affiliation{Stanford Institute for Materials and Energy Sciences, SLAC National Accelerator Laboratory and Stanford University, Menlo Park, California 94025, USA}
\affiliation{Department of Materials Science and Engineering, Stanford University, Stanford, CA 94305 USA}
\date{\today}

\begin{abstract}
Collective excitations contain rich information about photoinduced transient states in correlated systems. In a Mott insulator, charge degrees of freedom are frozen, but can be activated by photodoping. The energy-momentum distribution of the charge excitation spectrum reflects the propagation of charge degrees of freedom, and provides information about the interplay among various intertwined instabilities on the time scale set by the pump. To reveal charge excitations out of equilibrium, we simulate time-resolved x-ray absorption and resonant inelastic x-ray scattering using a Hubbard model. After pumping, the former resolves photodoping, while the latter characterizes the formation, dispersion, weight, and nonlinear effects of collective excitations. Intermediate-state information from time-resolved resonant inelastic x-ray scattering (trRIXS) can be used to decipher the origin of these excitations, including bimagnons, Mott-gap excitations, doublon and single-electron in-gap states, and anti-Stokes relaxation during an ultrafast pump. This paper provides a theoretical foundation for existing and future trRIXS experiments.
\end{abstract}
\maketitle

\section{Introduction}
The study of materials using ultrafast, pump-probe techniques recently has attracted considerable attention due to the ability to access novel states of matter, which can be tuned through changes in pump-pulse configurations\,\cite{zhang2014dynamics, basov2017towards}. State-of-the-art pump-probe techniques have pushed the temporal-resolution to pico- or even femtoseconds, enabling the observation of Floquet states\,\cite{wang2013observation, mahmood2016selective}, and transient superconductivity\,\cite{fausti2011light, mitrano2016possible} as well as the manipulation of both the amplitude and phase of various order parameters\,\cite{schmitt2008transient,boschini2018collapse,zong2019evidence}. Complementary to the control of matter at ultrashort timescales, the study of nonequilibrium states has also motivated ultrafast characterization of emergent physics through spectroscopic probes\,\cite{wang2018theoretical, buzzi2018probing}.  Time-resolved infrared spectroscopies and Raman scattering, provide information about particle-hole excitations near the Brillouin zone center\,\cite{rosker1986femtosecond, brorson1990femtosecond,saichu2009two,batignani2015probing,bowlan2018using}; and time-resolved angle-resolved photoemission (trARPES) can be used to obtain momentum-resolved single-particle spectra\,\cite{perfetti2006time, perfetti2008femtosecond}, with dynamical correlations inferred from the self-energy. The need to understand propagation of transient multi-particle excitations, \emph{e.g.} spin and charge fluctuations in unconventional superconductors\,\cite{kivelson2003detect}, frustrated magnets \cite{balents2010spin} and other materials with emergent phases or metal-insulator transitions\,\cite{imada1998metal}, necessitates resolving finite-momentum excitations out of equilibrium\,\cite{abbamonte2004imagin,reed2010effective, abbamonte2010ultrafast, wang2014real, wang2017producing, paeckel2019detecting}. To this end, a time-resolved technique recently was developed for resonant inelastic x-ray scattering (RIXS)\,\cite{dean2016ultrafast,mitrano2019ultrafast, parchenko2019orbital, mitrano2019evidence}.

Theoretical interpretation of the RIXS spectrum is complex, even in equilibrium, given the access to various x-ray edges and the tunability of light polarization.  In principle, RIXS can be used to detect a variety of collective excitations, including charge\,\cite{hill1998resonant, abbamonte1999resonant}, spin\,\cite{braicovich2009dispersion, le2011intense, dean2013persistence}, lattice\,\cite{chaix2017dispersive} and orbital\,\cite{schlappa2012spin,sala2014cairo} excitations, as well as hidden orders\,\cite{wray2015spectroscopic}. Though experimentally powerful, the theoretical elucidation of RIXS spectra usually relies on many-body numerical analysis due to the association with highly-excited states and strong correlations. Moreover, the potential for these collective excitations to overlap in energy requires unbiased theory and computation using microscopic models that explicitly include the relevant degrees of freedom\,\cite{ament2011resonant}. While much progress has been made over the past decade, the theoretical interpretation and numerical prediction of RIXS spectra remain a crucial task in quantum materials and photon science\cite{wang2018theoretical}.

Out of equilibrium, the physics revealed by time-resolved RIXS (trRIXS) can be even richer\,\cite{cao2019ultrafast}. Experimentally, Dean \emph{et al.}~successfully resolved fluctuating magnetic correlations in a pumped spin-orbit Mott insulator Sr$_2$IrO$_4$\,\cite{dean2016ultrafast}; and Mitrano \emph{et al.}~exploited trRIXS to characterize photoinduced decoherence of charge excitations in stripe-ordered La$_{2–x}$Ba$_x$CuO$_4$\cite{mitrano2019ultrafast,mitrano2019evidence}, where transient superconductivity had been observed in a previous study\,\cite{nicoletti2014optically}. Both experiments shed light on engineering and characterizing nonequilibrium states of matter.  Theoretically, a cross-section for trRIXS recently was derived from the light-matter interaction, specifically for the direct RIXS process, but neglecting both correlations in the valence shell and those induced by the core hole\,\cite{chen2019theory}. However, trRIXS should possess some distinct advantages in correlated systems, especially for understanding collective spin and charge excitations, where intermediate-state information, deciphered from the connection to time-resolved x-ray absorption spectrum (trXAS), can be used to disentangle the spectroscopic data. This necessitates a theoretical investigation of trRIXS spectra beyond simple noninteracting electrons.

\begin{figure*}[!t]
\begin{center}
\includegraphics[width=17cm]{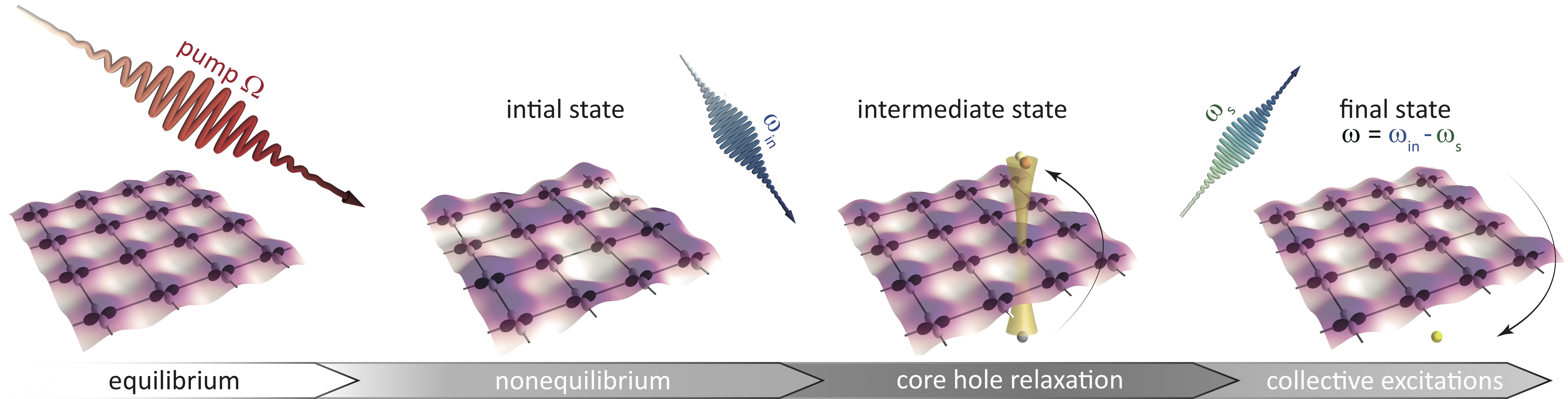}
\caption{\label{fig:cartoon} Schematic illustrating the pump-probe processes of indirect trXAS and trRIXS. Left to right: The pump field drives the electronic system out of equilibrium. After a certain time delay, the probe photon with a frequency $\win$ (resonant to an x-ray edge) excites a core-level electron into a high unoccupied level, leading to the trXAS spectrum. The core-hole attraction then generates collective excitations of the valence electrons. Finally, the photoelectron and core hole annihilate, emitting a photon with a frequency of $\wout$. This photon-in-photon-out process gives the trRIXS spectrum, which characterizes the remaining final state of the valence electrons, carrying collective excitations with energy $\win-\wout$. 
}
\end{center}
\end{figure*}

For strongly correlated systems, one typical example is the Hubbard model, whose Hamiltonian will be shown later in Eq.~\eqref{eq:Hubbard}. When the on-site Coulomb interaction $U$ is large, the ground state of this correlated system is known as a Mott insulating phase, which cannot be obtained perturbatively in a weak-coupling framework. On top of it, the doped Mott insulator is less understood. Intuitively, one may expect the localized electrons getting released by the doped carriers, together with the unraveling of correlation effects.\cite{lee2006doping} This process is reflected in the ARPES spectrum as the transfer of spectral weight from both lower- and upper-Hubbard bands to the quasiparticles near the chemical potential\cite{eskes1991anomalous}; and it is also reflected in $L$-edge RIXS spectrum as the damping of the magnons into paramagnons and the loss of the spin order\cite{le2011intense, dean2013persistence}. In contrast, the doping evolution of charge excitations is less understood. A key insight to be determined is how far these Mott-insulating features, especially the Mott gap, persist upon doping. In addition, can we expect the charge excitations gradually to evolve or suddenly switch into the particle-hole continuum, describable by a weak-coupling picture, at some particular doping? Does the transition occur through a closure of the Mott gap or a spectral weight transfer between different features? Due to the prevalent phases in cuprates emergent from the doped Mott insulator and their competition with the spin- and charge-ordered phases, the evolution of the charge and spin excitations needs to be better understood, and if possible, quantified. This necessity also extends to other strongly correlated materials such as iridates and nickelates\,\cite{ishii2011momentum, hepting2019electronic}, where various phases are also believed to be emergent from the doped Mott insulator.

To answer the above questions, an ideal approach is to continuously dope the materials off the parent compound without being affected by any uncontrollable factors resulting from chemical substitution or sample quality. One promising realization is through the field-effect gating, which could continuously tune carrier density in the ultrathin cuprate films\,\cite{bollinger2011superconductor,ahn1999electrostatic}. This technique is currently restricted to the thickness of the sample. An alternative approach is through the photodoping induced by an ultrafast laser pump.\cite{rameau2014photoinduced} The trRIXS technique then allows us to characterize the evolution of charge and spin excitations in this ultrafast process. In addition to the characterization of continuous doping dependence, special pump conditions may induce novel states beyond the equilibrium phase diagram of a doped Mott insulator, such as the transient superconducting phases above $T_c$\,\cite{fausti2011light}.

For this purpose, we report a numerical study of trXAS and trRIXS in strongly correlated systems, using the two-dimensional single-band Hubbard model, as a proxy for cuprate superconductors.  We provide a picture of possible low-energy photoinduced states and excitations, where trXAS captures photoinduced many-body states and characterizes photodoping, while \emph{indirect} trRIXS tracks the evolution of Mott-gap excitations, bimagnon excitations, and in-gap excitations induced by photodoping. In addition to energy renormalization and redistribution of spectral weight, which are accessible in optics, trRIXS provides a unique characterization of these \emph{dispersive} transient excitations. Moreover, trRIXS allows us to dissect the photoinduced excitations through the intermediate states. With the tunability of the incident energy through an x-ray edge, trRIXS spectra provide a direct and continuous visualization of the insulator-metal transition induced by photodoping.

The organization of this paper is as follows. We first introduce the theory for trXAS and trRIXS in Sec.~\ref{sec:RIXStheory}, with a focus on indirect x-ray edges (core-to-above-the-valence-shell excitations). We apply this theory to the Hubbard model, as a simple prototype for correlated materials, and present the trXAS and trRIXS results in Sec.~\ref{sec:numerics}. We specifically analyze the momentum dependence and the impact of the core-hole lifetime and the probe pulse in Sec.~\ref{sec:discussion}. Finally, we conclude our study in Sec.~\ref{sec:summary}, with  an outlook for trRIXS simulations of other correlated systems.

\section{Pump-Probe Theory for XAS and RIXS}\label{sec:RIXStheory}

In this section, we introduce the spectroscopic theory for trXAS and trRIXS. Most of the derivations follow Ref.~\onlinecite{chen2019theory}, extending the derivations to indirect processes for correlated systems.

In second quantization, the light-electron interaction can be described through a Peierls substitution $c_{\mathbf{i}\sigma} \rightarrow c_{\mathbf{i}\sigma} \exp[-i\int_{-\infty}^{\textbf{r}_i} \textbf{A}(\rbf^\prime,t)\cdot d\mathbf{r}^\prime] $. Here, $c_{\mathbf{i}\sigma}$ is the annihilation operator for an electron with spin $\sigma$ on site $\mathbf{i}$, and $\textbf{A}$ is a vector potential of the light. In a pump-probe process, this is the full vector potential, accounting for both the pump and probe light, which we denote as $\textbf{A}^{\rm(pump)}$ and $\textbf{A}^{\rm(pr)}$, respectively. Since the probe field is typically much weaker than the pump ($|\textbf{A}^{\rm(pr)}| \ll |\textbf{A}^{\rm(pump)}|$), one can expand the pump-probe Hamiltonian perturbatively in the probe field
\begin{eqnarray}
\Ham(t) = \Ham_0(t) + \Ham_{\rm pr}(t)\,.
\end{eqnarray}
The $\Ham_0(t)$ contains the Hamiltonian with the presence of a pump field $\mathbf{A}^{\rm(pump)}(t)$; the $\Ham_{\rm pr}(t)$ contains the part induced by the probe field, typically to second order,
\begin{eqnarray}\label{chp2:eleLightHam}
\mathcal{H}_{\rm pr}(t) &=& -\sum_{\qbf,\eps} j_{\eps}(\qbf,t) \probA_\eps(\qbf,t)\nonumber\\
&&-\frac12\! \sum_{\qbf,\qbf_i\atop\epsin,\epsout} \gamma_{\epsin\epsout}(\qbf,t)\probA_\epsout(\qbf_s,t)^*\probA_{\epsin}(\qbf_i,t),
\end{eqnarray}
where $\probA_\eps(\qbf,t) \!=\! \int e^{-i\qbf\cdot \rbf^\prime}\!\textbf{A}^{\rm(pr)}(t)\!\cdot\! \mathbf{e}_\eps dr^\prime$, and $\eps$ denotes the light polarization direction of the probe field. ${j}_{\eps}$ and $\gamma_{\epsin\epsout}$ are respectively the current density and scattering vertex operator with a momentum transfer $\qbf\!=\!\qin\!-\!\qout$. 

To describe resonant absorption, we adopt the second quantized form of the photon field, $\probA_\eps(\qbf) = a_{\qbf\eps}+a_{-\qbf\bm{\varepsilon}}^\dagger$. Expanding $\Ham_{\rm pr}$ to second order in $\probA_\eps$ and with momentum conservation,
\begin{eqnarray}\label{eq:probeHamDiv}
\Ham_{\rm pr}(t) = \mathcal{H}_{\rm pr}^{\rm (ab)}(t) + \mathcal{H}_{\rm pr}^{\rm (ab2)}(t) + \mathcal{H}_{\rm pr}^{\rm (sc)}(t) +h.c.
\end{eqnarray}
where single-photon absorption, two-photon absorption, and scattering are\,\cite{wang2018theory, chen2019theory}
\begin{widetext}
\begin{eqnarray}\label{eq:generalAbsEq}
   \left\{
   \begin{array}{l}
       \mathcal{H}_{\rm pr}^{\rm (ab)}(t)  = -\sum_{\alpha,\beta}\sum_{\qbf\eps}\sum_{\kbf\sigma} M^{(1)}_{\alpha\beta}(\qbf, \kbf,t) c_{\kbf+\qbf,\sigma}^{(\alpha)\dagger} c_{\kbf\sigma}^{(\beta)}  a_{\qbf\eps},\\
       \mathcal{H}_{\rm pr}^{\rm (ab2)}(t)  = -\frac12 \sum_{\alpha,\beta}\sum_{\qbf_i\qbf_s\atop\eps_i,\eps_s}\sum_{\kbf\sigma} M^{(2)}_{\alpha\beta}(\qbf_i-\qbf_s, \kbf,t) c_{\kbf+\qbf,\sigma}^{(\alpha)\dagger} c_{\kbf\sigma}^{(\beta)}  a_{-\qbf_s\eps}a_{\qbf_i\eps^\prime},\\
       \mathcal{H}_{\rm pr}^{\rm (sc)}(t)  = -\frac12 \sum_{\alpha,\beta}\sum_{\qbf_i\qbf_s\atop\eps_i,\eps_s}\sum_{\kbf\sigma} M^{(1)\prime}_{\alpha\beta}(\qbf_i-\qbf_s, \kbf,t) c_{\kbf+\qbf,\sigma}^{(\alpha)\dagger} c_{\kbf\sigma}^{(\beta)}a_{\qbf_s\eps}^\dagger a_{\qbf_i\eps^\prime}.
   \end{array}
    \right.
\end{eqnarray}
The matrix elements $M^{(1)}_{\alpha\beta}$, $M^{(2)}_{\alpha\beta}$, and $M^{(1)\prime}_{\alpha\beta}$ are determined by the transition rate between different orbitals.

Using $\Ham_{\rm pr}$ in the interaction picture, we can expand the time-evolution operator $\mathcal{U}(t,-\infty)$ to second order in the probe field
\begin{eqnarray}\label{eq:Uexpr}
\mathcal{U}(t,-\infty) &\approx& \mathcal{U}_0(t,-\infty)- i\int_{-\infty}^t \hat{\Ham}_{\rm pr}^{\rm (ab)}(\tau)  d\tau    - i\int_{-\infty}^t\!dt_2 \int_{-\infty}^{t_2}\!dt_1\,  \hat{\Ham}_{\rm pr}^{\rm (ab)}(t_2)  \hat{\Ham}_{\rm pr}^{\rm (ab)}(t_1) \nonumber\\
&& - i\int_{-\infty}^tdt_2 \int_{-\infty}^{t_2}dt_1\, \hat{\Ham}_{\rm pr}^{\rm (ab)\dagger}(t_2)   \hat{\Ham}_{\rm pr}^{\rm (ab)}(t_1) - i\int_{-\infty}^t \hat{\Ham}_{\rm pr}^{\rm (ab2)}(\tau)  d\tau- i\int_{-\infty}^t \hat{\Ham}_{\rm pr}^{\rm (sc)}(\tau)  d\tau.
\end{eqnarray}
\end{widetext}
Here, $\mathcal{U}_0(t_2,t_1) = \mathcal{T}e^{-i\int_{t_1}^{t_2} \!\Ham_0(\tau) d\tau}$, where $\mathcal{T}$ is the time-ordering operator, such that in the interaction picture any operator $\hat{O}(t) = \mathcal{U}_0(-\infty,t)O(t)\mathcal{U}_0(t,-\infty)$.
Since $| \psi(t = -\infty) \rangle$ is usually the equilibrium ground state when the temperature is zero and the optical pump is off-resonance 
from x-ray edges, the Hermitian conjugate terms of $\mathcal{H}_{\rm pr}^{\rm (ab)}$ and $\mathcal{H}_{\rm pr}^{\rm (ab2)}$ do not contribute to the first four integrals, because the ground state cannot emit any photons. For convenience, we use $\langle \cdots\rangle_0$ to denote the expectation under $| \psi(t = -\infty) \rangle$.

To capture both the electronic system and the quantized photon field, we write down the wavefunction as $|\psi(t)\rangle = |\Psi(t)\rangle_e\otimes |N(t)\rangle_{\rm ph}$, where the incident photon state is a coherent state
\begin{eqnarray}\label{eq:photonfield}
    |N(t)\rangle_{\rm ph} = e^{-N(t)/2}e^{\sqrt{N(t)}a_{\qin\epsin}^\dagger}|0\rangle.
\end{eqnarray}
The average photon number $N(t)$ semi-classically describes the profile of the time-dependent probe field\,\cite{wang2018theory, chen2019theory}. We can define a probe shape function $g(\tau;t) = \sqrt{N(\tau-t)}$ for a probe pulse centered at time $t$. Since the bare probe photon propagator describes non-interacting photons with an energy $\win$, the annihilation operator gives
\begin{eqnarray}\label{eq:coherentState}
a_{\qin\epsin}|\psi(\tau)\rangle  = e^{-i\win \tau}g(\tau;t)|\psi(\tau)\rangle.
\end{eqnarray}
In this paper, we approximate the probe-shape function $g(\tau;t)$ by a Gaussian pulse with width $\sigma_{\rm pr}$
\begin{eqnarray}\label{eq:probeProfile}
g(\tau;t) = \frac1{\sqrt{2\pi}\sigma_{\rm pr}} e^{-(\tau-t)^2/2\sigma_{\rm pr}^2}\,,
\end{eqnarray}
though the precise temporal profile can be determined more directly in experiment.

By selecting the probe photon field and the observable $\mathcal{O}$, one specifies a distinct process through the expectation value $\langle \mathcal{O}(t)\rangle$. Measuring the difference of the photon number with the incident frequency, the first-order response gives photon absorption\,\cite{chen2019theory}. For a resonant x-ray probe, this probe ``current'' can be replaced by a dipole transition operator ${j}_{\eps}(\qbf) \propto \mathcal{D}_{\qbf}$ due to the dominant inter-band transition process. For an indirect x-ray edge, \emph{e.g.}~Cu $K$-edge, we simplify the dipole operator as (see the schematic in Fig.~\ref{fig:cartoon})
\begin{eqnarray}
\mathcal{D}_{\qbf} =\frac1{\sqrt{N}}\sum_{\ibf,\sigma} e^{i\qbf\cdot\rbf_\ibf} h_{\ibf,\sigma}^\dagger d_{\ibf,\sigma},
\end{eqnarray}
which excites a core-level electron (annihilation operator $d_\ibf$) to a high-energy unoccupied level (annihilation operator $h_{\ibf}$).  We denote this edge energy in the atomic limit as $E_{\rm edge}$. Since both the core-level and high-energy states lie far from the valence and conduction bands, we ignore the matrix element in this case. The impact of the resonant excitation, in both indirect XAS and RIXS, is reflected only through the attraction between valence electrons and the core hole [defined later in Eq.~\eqref{eq:Hubbard}].

The trXAS spectrum measures the difference between the outgoing and incoming photon number, which is dominated by the second term in Eq.~\eqref{eq:Uexpr}, as
\begin{eqnarray}\label{eq:generalAbsEq}
    \iint_{-\infty}^\timeMax dt_1dt_2\langle \hat{\Ham}_{\rm pr}^{\rm (ab)\dagger}(t_2)\hat{a}_{\eps}^\dagger(\timeMax) \hat{a}_{\eps}(\timeMax)\hat{\Ham}_{\rm pr}^{\rm (ab)}(t_1)\rangle_0
    \end{eqnarray}
Evaluating the photon operators on the coherent state through Eq.~\eqref{eq:coherentState}, we obtain the leading-order absorption ratio\,\cite{shvaika2012exact}
\begin{eqnarray}\label{eq:XASexpr}
    \mathcal{B}(\win, t)& =& \iint_{-\infty}^\timeMax g(t_1;t)g(t_2;t) e^{i\win(t_2-t_1)}\nonumber\\
    &&\langle \hat{\mathcal{D}}_{\qbf\eps}^\dagger(t_2) \hat{\mathcal{D}}_{\qbf\eps}(t_1)\rangle_0 dt_1dt_2
\end{eqnarray}
In practice, $\timeMax$ can be taken to $+\infty$ when the detection time is much longer than the probe pulse. 
Due to the locality of the core-level electronic wavefunction, we usually assume the core-hole induced by the dipole transition is immobile. In this sense, the sequential dipole operators have an implicit spatial relationship
\begin{eqnarray}\label{eq:coreholeReduction}
 \mathcal{D}_{\qbf_2}^\dagger \mathcal{D}_{\qbf_1} = \sum_{\ibf ,\sigma';\jbf,\sigma}  d^{\dagger}_{\jbf\sigma}h_{\jbf\sigma} h^\dagger_{\ibf\sigma^\prime} d_{\ibf\sigma^\prime} e^{i(\qbf_1-\qbf_2)\cdot \rbf_\ibf}\delta_{\ibf\jbf}.
\end{eqnarray}
Therefore, the XAS cross-section provides effectively no $\qbf$ information.

When the incident photon is off-resonant with any intermediate states, the photon absorption or scattering occurs via a virtual state, and in this case the non-resonant Raman contributions dominate in the cross-section\,\cite{devereaux2007inelastic,wang2018theory}. Instead, the resonant part dominates for a probe at an x-ray edge. 
Therefore, the fourth term of Eq.~\eqref{eq:Uexpr} involving intermediate states can be interpreted as trRIXS where the scattered photon intensity 
\begin{eqnarray}\label{eq:RIXScrossSec1}
\langle n^{\rm ph}_{\qbf_s}\rangle \!&=&\! \iint_{-\infty}^\timeMax\! \int_{-\infty}^{t_2^\prime}\int_{-\infty}^{t_1^\prime}\!\big\langle \hat{\Ham}_{\rm pr}^{\rm (ab)\dagger}(t_2)\hat{\Ham}_{\rm pr}^{\rm (ab)}(t_2^\prime)  \hat{a}^\dagger_{\qbf_s\eps}(\timeMax)\nonumber\\
&&\mkern-18mu\hat{a}_{\qbf_s\eps}(\timeMax)\hat{\Ham}_{\rm pr}^{\rm (ab)\dagger}(t_1^\prime)\hat{\Ham}_{\rm pr}^{\rm (ab)}(t_1)  \big\rangle_0 dt_1 dt_2 dt_1^\prime dt_2^\prime .
\end{eqnarray}
Note that the probe Hamiltonian is explicitly time dependent due to the presence of the non-perturbative pump field. That being said, the dipole transition operator $\hat{\mathcal{D}}_{\qbf}(t)$, which will appear later in Eq.~\eqref{eq:RIXScrossSec3}, is affected by the Peierls substitution of $\mathbf{A}^{\rm(pump)}(t)$. However, since we have employed a sudden approximation while simplifying the light-matter interaction via a dipole transition, such a gauge shift cancels between the initial and final electronic state of the transition, leaving a bare momentum transfer $\qbf$.

Employing the coherent photon state, we can simplify the photon expectation value into a single term
\begin{eqnarray}\label{eq:normalOrderExp4PhotonRes}
&&\mkern-18mu\langle \hat{a}_{\qbf_{\rm i}\eps_{\rm i}}^\dagger(t_2) \hat{a}_{\qbf_2\eps_2}(t_2^\prime) \hat{a}_{\qbf_{\rm s}\eps_{\rm s}}^\dagger(\timeMax) \hat{a}_{\qbf_{\rm s}\eps_{\rm s}}(\timeMax) \hat{a}_{\qbf_1\eps_1}^\dagger(t_1^\prime) \hat{a}_{\qbf_{\rm i}\eps_{\rm i}}(t_1)\rangle_0\nonumber\\
&&\mkern54mu\approx e^{i\win(t_2-t_1)-i\wout(t_2^\prime-t_1^\prime)} g(t_1,t)g(t_1^\prime,t) \nonumber \\
&&\mkern198mu\delta_{\qbf_{\rm s}\qbf_2}\delta_{\eps_{\rm s}\eps_2}\delta_{\qbf_{\rm s}\qbf_1}\delta_{\eps_{\rm s}\eps_1}.
\end{eqnarray}
because the scattered photon intensity is much smaller than the incident photon intensity.  As shown in Fig.~\ref{fig:cartoon}, this cross-section involves a two-step photon-in-photon-out process, leaving the electronic system in an excited state, characterized by various collective excitations.

Beyond Eq.~\eqref{eq:RIXScrossSec1}, the empirical trRIXS cross-section depends on other factors outside the present theory. First, the lifetime of the intermediate state is comparable to or smaller than the probe timescale and must be explicitly taken into account when modeling the cross-section. This spontaneous irreversible decay process can be included phenomenologically by replacing $\mathcal{U}_0(t_1^\prime,t_1)$ with $\mathcal{U}_0(t_1^\prime,t_1)l(t_1^\prime-t_1)$, where $l(\tau)\! =\!  e^{- \tau/\tau_{\rm core}}\theta(\tau)$ describes the decay of the core hole with $\tau_{\rm core}$ as the core hole lifetime. Second, assuming a local core hole, \emph{i.e.}~Eq.~\eqref{eq:coreholeReduction}, the sequential dipole operator $\mathcal{D}_{\qout}^\dagger\mathcal{D}_{\qin}$ appearing in the cross-section becomes a function of only the momentum transfer $\qbf = \qin-\qout$. Finally, the role of light polarization is effectively irrelevant for the indirect process, due to the lack of spin-orbit coupling\,\cite{jia2012uncovering,jia2014persistent}.

In the end, the trRIXS cross-section can be written as
\begin{eqnarray}\label{eq:RIXScrossSec3}
\mathcal{I}(\qbf,\omega,\win,t)\!&\mkern-6mu=&\!\mkern-12mu\iiiint
e^{i\win(t_2-t_1)-i\wout(t_2^\prime-t_1^\prime)} 
  g(t_1,t)g(t_2,t) \nonumber\\
&&\mkern-42mu\times \big\langle \hat{\mathcal{D}}_{\qin\epsin}^\dagger(t_2)\hat{\mathcal{D}}_{\qout\epsout}(t_2^\prime) \hat{\mathcal{D}}_{\qout\epsout}^\dagger(t_1^\prime)   \hat{\mathcal{D}}_{\qin\epsin}(t_1)  \big\rangle_0\nonumber\\
&&\mkern-42mu\times l(t_1^\prime-t_1)l(t_2^\prime-t_2) dt_1  dt_2 dt_2^\prime dt_1^\prime\,.  
\end{eqnarray}
In numerical calculations, we typically evaluate the cross-section by discrete integration, where the number of discrete time steps is $N_t$. Here, we want to emphasize that the pump-probe spectral calculation for trARPES\cite{freericks2009theoretical}, nonequilibrium spin and charge structure factors\cite{wang2017producing}, non-resonant Raman scattering\cite{wang2018theory}, and XAS [as in Eq.~\eqref{eq:XASexpr}] would require knowledge of the correlator in the integrand at time pairs $(t_1,t_2)$ in the 2D time plane with a computational complexity $O(N_t^2)$. However, the quadruple integral in Eq.~\eqref{eq:RIXScrossSec3} requires evaluation in a 4D time hyperplane, such that the trRIXS calculation would have a time complexity $O(N_t^4)$, though some appropriate truncation due to finite core-hole lifetime and probe width could reduce the computational cost. In general, trRIXS also requires a traversal of $(\qin,\qout)$ pairs. However, this momentum traversal can be reduced to a sequence of the difference $\qin-\qout$, when core hole is local as mentioned above.

\section{Numerical Results}\label{sec:numerics}
We focus on indirect x-ray spectra in strongly correlated systems, specifically taking the Cu $K$-edge in cuprates as an example. We consider valence and conduction electrons describable by the 2D single-band Hubbard model, with an additional core-hole (attractive) potential, given by
\begin{eqnarray} \label{eq:Hubbard}
\mathcal{H}_0& =& -\sum_{\ibf,\jbf,\sigma}\!t_h^{\ibf\jbf} ( c_{\jbf\sigma}^\dagger c_{\ibf\sigma}\!+\!h.c.)\! + U\!\sum_\ibf \!n_{\ibf\uparrow} n_{\ibf\downarrow}\nonumber\\
&& + E_{\rm edge}\sum_\ibf n^d_{\ibf} - U_c\!\sum_\ibf \!n_{\ibf} n^d_{\ibf},
\end{eqnarray}
where $n^d_{\ibf}\!=\!d_\ibf d_\ibf^\dagger$ is the core-hole number operator. We truncate the kinetic energy to the nearest-neighbor $t_h$ and next-nearest-neighbor $t^\prime_h$ hopping in a tight-binding picture and set $t^\prime_h \!=\!-0.3\,t_h$ and $U\!=\!8t_h$. These model parameters have been widely used to describe the bare bandstructure for cuprates with particle-hole asymmetry, where the nearest-neighbor hopping $t_h$ is typically $\sim300$\,meV. The above parameter choice leads to the spin-exchange energy $J=4\,t_h^2/U=0.5t_h\sim 150$\,meV. At the Cu $K$-edge, the edge energy $E_{\rm edge}\sim 8.98$keV, which is selected as the baseline for $\win$. The core-hole potential $U_c$ is set as 12$t_h$. Due to the $O(N_t^4)$ complexity of the trRIXS calculation, we adopt the 12D Betts cluster as a compromise between complexity and finite size. Throughout, we focus on the half-filled Hubbard model as a prototype of the Mott insulator. This system spans a $\sim 10^{6}$ many-body Hilbert space, accounting for the presence of the core hole, and one can reasonably evaluate the cross-section by considering $\sim 10^{11}$ points in the 4D time hyperplane.

We simulate the pump field as an oscillatory Gaussian vector potential 
\begin{equation}
    \mathbf{A}^{\rm(pump)}(t) = A_0\, e^{-t^2/2\sigma^2} \cos(\Omega t)\, \hat{\mathbf{e}}_{\rm pol}.
\end{equation}
Throughout this paper, we select the pump frequency $\Omega=5\,t_h$, which corresponds to the titanium-sapphire 800nm laser for $t_h=300$meV, intentionally selected to be above the Mott gap ($\sim\! 4-5\,t_h$ here for $U=8t_h$). We set the pump width as $\sigma=3\,t_h^{-1}$, corresponding to a $250$\,fs pulse (FWHM). To highlight pump-induced effects, we employ a strong pump field $A_0=0.4$, which gives a fluence of 24 mJ/cm$^2$\cite{fluence}. Due to the tilted geometry of the 12D Betts cluster, we choose $\hat{\mathbf{e}}_{\rm pol}$ diagonal in momentum space, which corresponds to a linear pump in the $ac$ plane in real space. In a typical experiment, $\tau_{\rm core}$ is finite, but much smaller than the probe width. Thus, we set the probe width $\sigma_{\rm pr}\! =\! 1.5\,t_h^{-1}$ and $\tau_{\rm core}\! =\! 0.5\,t_h^{-1}$ in the following calculations.

In this section, we present the numerical calculations of trXAS and trRIXS, with the purpose of characterizing the photodoping and collective excitations. We use the parallel Arnoldi method\,\cite{lehoucq1998arpack,  jia2017paradeisos} to determine the equilibrium ground-state wavefunction, and the Krylov subspace technique\,\cite{manmana2007strongly, balzer2012krylov} to evaluate the time-evolution. The calculation is performed at zero-temperature.

\begin{figure}[!t]
\begin{center}
\includegraphics[width=\columnwidth]{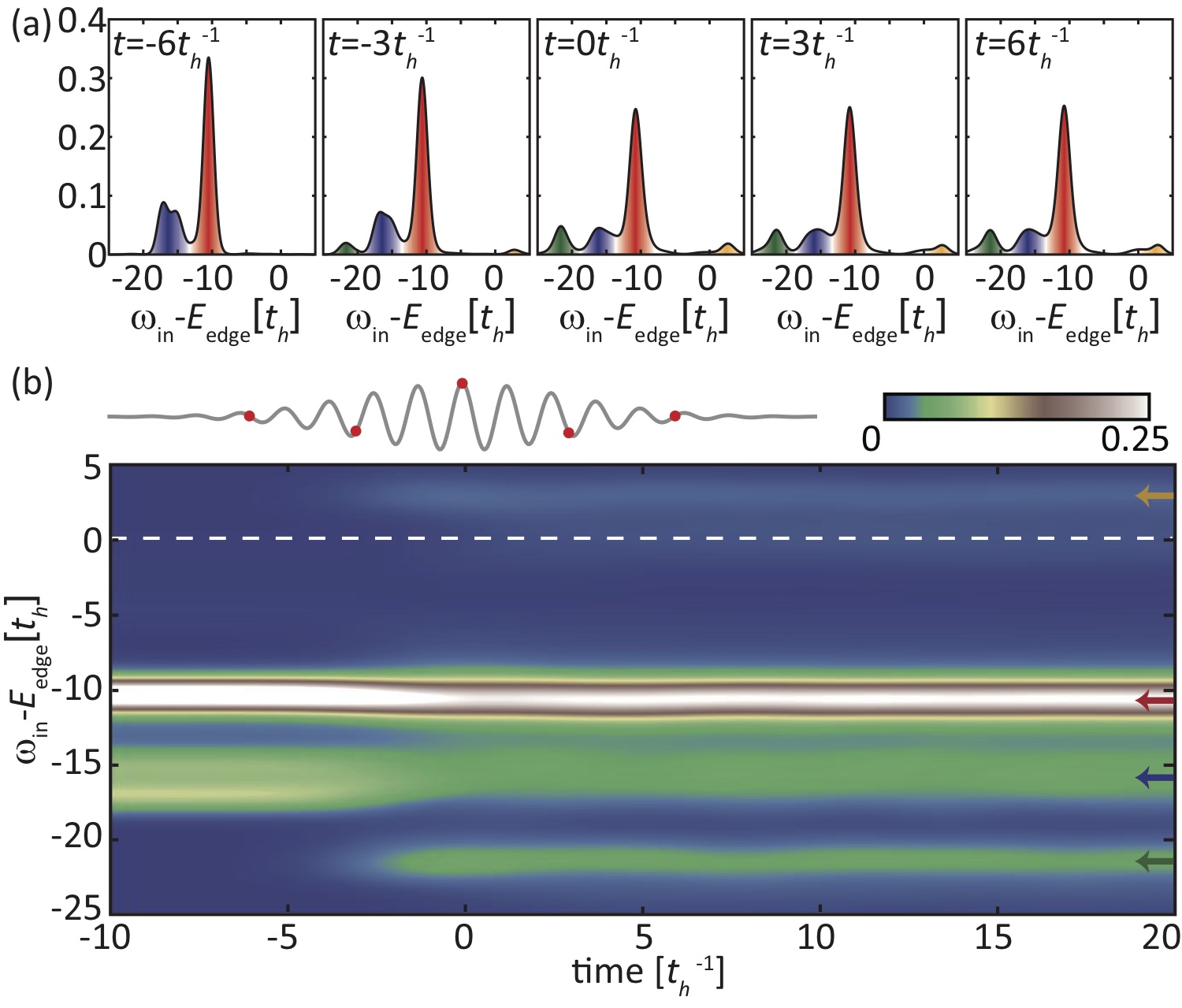}
\caption{\label{fig:xasSpec}(a) Snapshots of trXAS spectra at $t=-6t_h^{-1}$, $-3t_h^{-1}$, 0, $3t_h^{-1}$, and $6t_h^{-1}$. Different peaks are marked by different colors, corresponding to the processes in Fig.~\ref{fig:xasComp}. (b) False-color plot of the trXAS spectra during and after the pump. The pump pulse is indicated by the white curve and the times of the snapshots in panel (a) are marked by red circles. The colored arrows denote the different processes of panel (a).
}
\end{center}
\end{figure}

\subsection{trXAS and Photodoping}
\begin{figure}[!t]
\begin{center}
\includegraphics[width=\columnwidth]{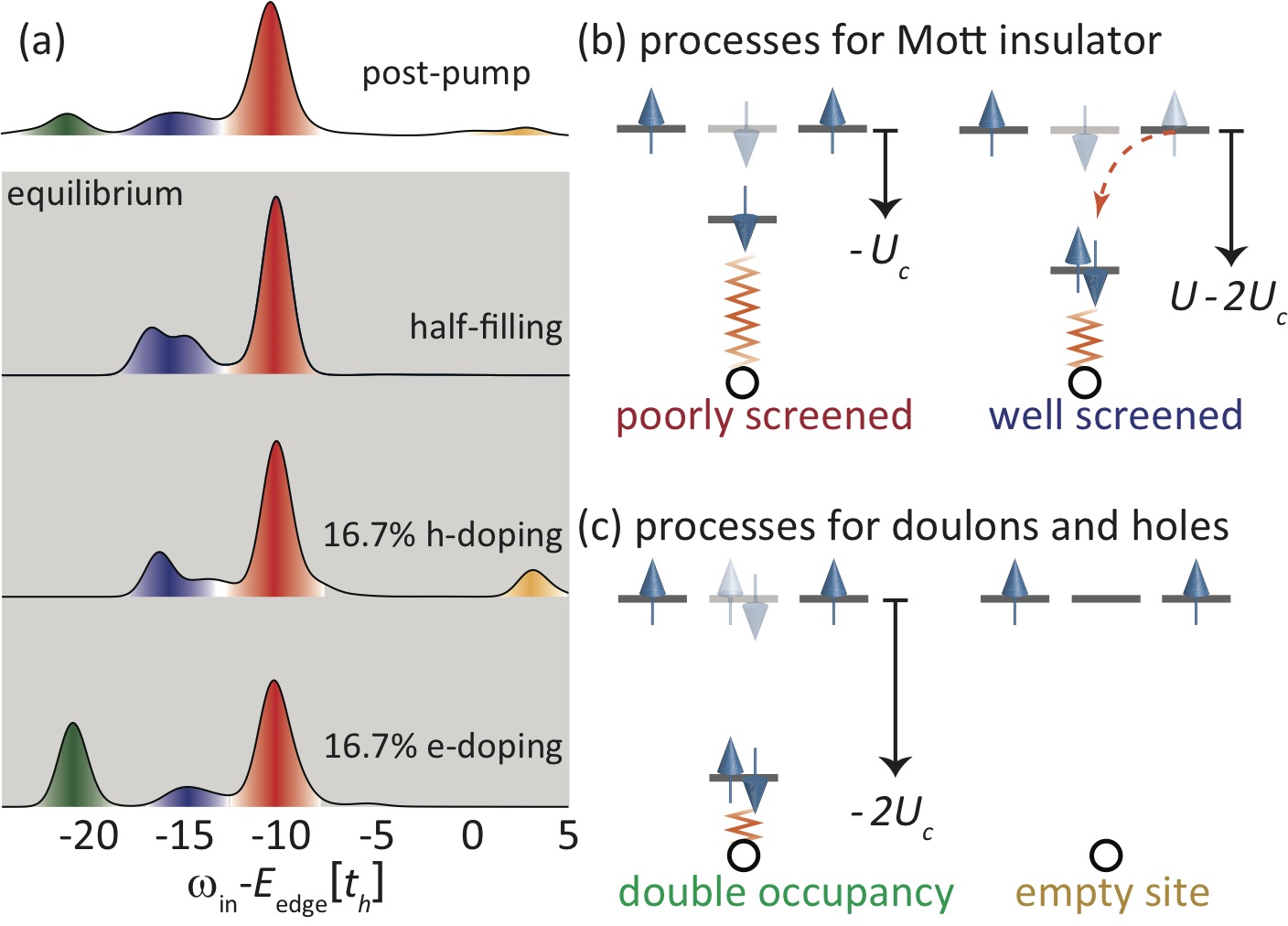}
\caption{\label{fig:xasComp}(a) Upper to lower sprectrum: Post-pump trXAS ($t\!=\!10t_h^{-1}$) for half-filled system in Fig.~\ref{fig:xasSpec}, equilibrium XAS for the same half-filled system, equilibrium XAS for 16.7\% hole-doped and 16.7\% electron-doped systems, respectively.
(b,c) Schematic of four types of resonant absorption processes with intermediate states: (b) the ``poorly-screened'' and ``well-screened'' channels present at half-filled Mott insulators; (c) processes corresponding to double occupancy (doublon) and empty site (hole).  The four resonances are denoted by different colors in (a), following the color code of (b,c).
}
\end{center}
\end{figure}

Figure~\ref{fig:xasSpec} shows the trXAS spectrum calculated with the above Hubbard model and linear pump conditions. Before pumping, the half-filled Hubbard model displays two absorption peaks [see the left panel of Fig.~\ref{fig:xasSpec}(a)]. These two absorption peaks, separated from each other due to the presence of strong correlations, are attributed to the so-called ``poorly-screened'' and ``well-screened'' states\,\cite{jia2012uncovering, kim2002resonant, lu2006incident}. As shown in the schematic of Fig.~\ref{fig:xasComp}(b), when the dipole transition occurs with the core hole on a singly occupied site, the system gains energy $\sim E_{\rm edge}-U_c$ (atomic limit), corresponding to the ``poorly-screened'' peak at $\win\approx E_{\rm edge}-10.5t_h$ (cluster calculation); on the other hand, when the core-hole attraction induced by the dipole transition induces a doublon by pulling an electron from a neighboring site, the system gains energy $\sim E_{\rm edge}+U-2U_c$, corresponding to the ``well-screened'' peak at $\win\approx E_{\rm edge}-16t_h$. Due to strong correlations, the ground state of the half-filled Hubbard model is dominated by singly occupied initial states, which lead to these two consequences during x-ray photoexcitation. The nature of these two absorption peaks has been well studied in equilibrium, both theoretically and experimentally\cite{jia2012uncovering, kim2002resonant, lu2006incident}.

After the pump, two extra absorption peaks at $\win\sim\! E_{\rm edge}-2U_c$ (green) and $\win\sim\! E_{\rm edge}$ (yellow) arise, which are absent in equilibrium [see Fig.~\ref{fig:xasSpec}(b) and right panels in Fig.~\ref{fig:xasSpec}(a)]. For clarity, we denote the original two peaks in equilibrium as the ``major peaks'' as their intensities are still dominant after the pump, while we denote these two extra peaks induced by the pump as ``minor peaks''. The intensity of these two minor peaks rises as that of the two major ones drops, due to the conservation of total spectral weight. Despite the variation of intensity, the energy position of all four peaks changes little with time. Due to the generation of doublons and holes and their impact on the single-particle kinetic energy, the two major peaks shift slightly, negligible when compared to the peak separation (on the scale of $U$ and $U_c$). These time-domain behaviors reflect the locality of the XAS excitations in an indirect process and indicate simple spectral weight transfer among various many-body states in the equilibrium manifold.

\begin{figure*}[!th]
\begin{center}
\includegraphics[width=18cm]{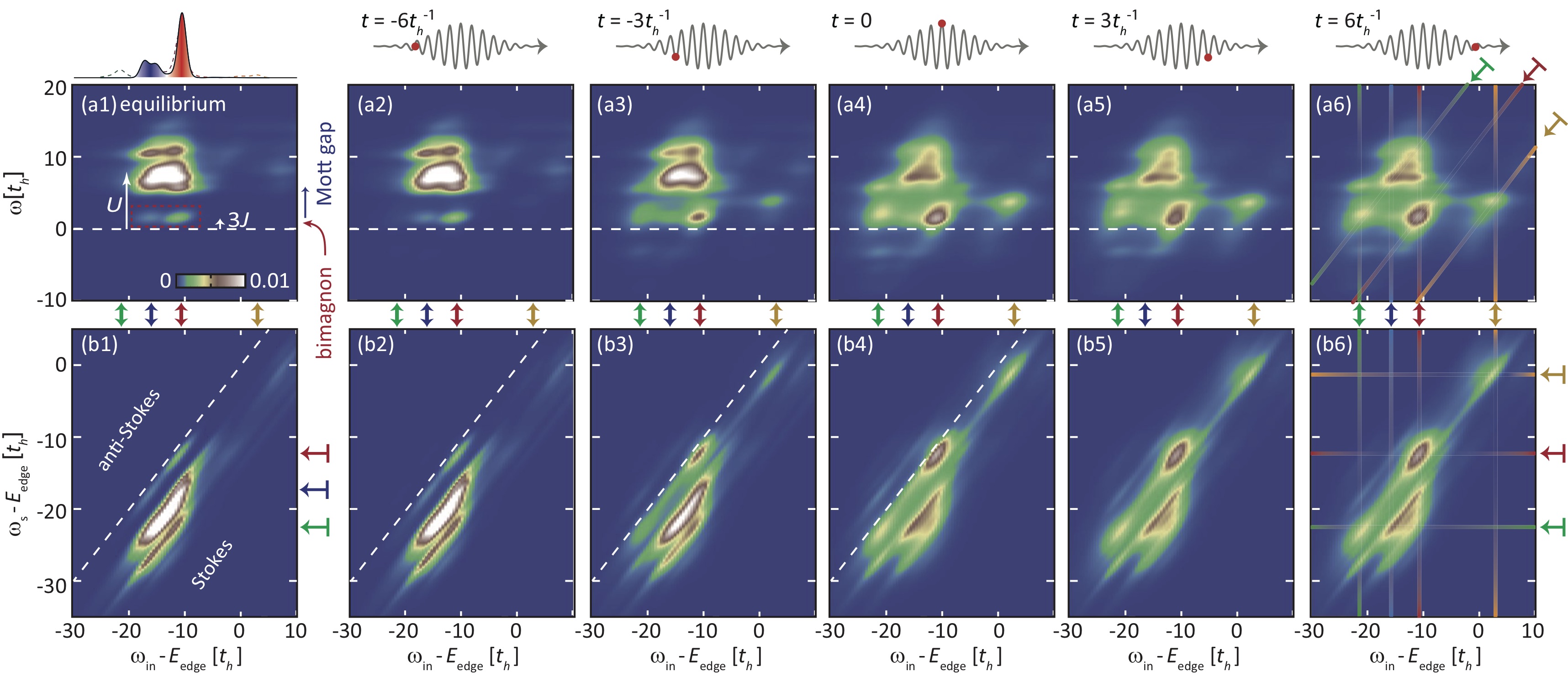}
\caption{\label{fig:rixs} The $\qbf=(\pi,\pi/3)$ trRIXS spectra during a $\Omega\!=\!5t_h$ pump in (a1-a6) energy loss-incident energy ($\omega-\win$) and (b1-b6) scattering energy-incident energy ($\wout-\win$) view. The four XAS resonances $\win-E_{\rm edge}$=\,-21.5, -16, -10.5, and 2.5\,$t_h$ are labeled as green, blue, red and yellow arrows, respectively, corresponding to the ``double-occupancy'', ``poorly-screened'', ``well-screened'' and ``empty-site'' intermediate states.
The arrows to the right denote the final-state resonances of $\wout=-1.32$ (yellow), -12.27$t_h$ (red), -17.62$t_h$ (blue) and -22.53$t_h$ (green). The white dashed lines denote the elastic response, while the colored lines in (a6) and (b6) represent the resonance for initial and final states.
The upper insets show the equilibrium XAS spectrum and the corresponding time during a pump pulse.
}
\end{center}
\end{figure*}

The origin of these photoinduced minor peaks can be revealed by comparing with the doped equilibrium systems. Figure \ref{fig:xasComp}(a) shows the comparison of the post-pump trXAS spectrum (for the half-filled system at $t=10\,t_h^{-1}$) and the equilibrium XAS spectra obtained for half-filled, 16.7\% hole-doped and 16.7\% electron-doped systems. Obviously, these minor peaks induced by the pump match the XAS features of the electron- or hole-doped systems, respectively. In a naive local picture as shown in Fig.~\ref{fig:xasComp}(c), the peak positions reflect the energy cost of a charge impurity in the presence of doublons or holes: A pre-existing doublon screens the core hole at a cost of $E_{\rm edge}-2U_c$ -- the so-called ``double-occupancy'' channel, while the holes are not affected by the core hole, contributing to a $\win\!\sim\!E_{\rm edge}$ hump. (Due to the presence of quantum fluctuations, the spectral peaks in the cluster calculation hardens by $\sim2.5t_h$ for both channels.) The coexistence of these two peaks in the trXAS reflects the post-pump state being a superposition of hole- and electron-doping. In contrast to trARPES, where doping-induced features are highly affected by quantum fluctuations\,\cite{wang2017producing}, the trXAS features are well-separated in energy and are sensitive to local configurations, with relatively rigid peak positions. In this sense, the trXAS spectrum provides a better parametrization of the overall electronic configuration after the pump. This is important for the ultrafast control of effective doping, especially for those materials where a wide range of doping is not easily accessible through equilibrium chemical synthesis \emph{e.g.} Sr$_2$IrO$_4$, LaNiO$_3$ and Nd$_2$CuO$_4$\,\cite{armitage2010progress, scherwitzl2011metal,cao2016hallmarks}.

\subsection{trRIXS and Collective Excitations}\label{sec:trRIXSres}
The photodoping characterized by trXAS enables further study about the dynamics of elementary excitations associated with a photodoped Mott insulator.
This can be accessed by trRIXS. In this section, we first focus on a momentum transfer $\qbf\!=\!(\pi,\pi/3)$ near the anti-ferromagnetic zone boundary (AFZB). The selection of momentum is due to the fact that the bimagnon excitation (introduced later in this section) is found most prominently near the antinode in the Cu $K$-edge RIXS of La$_2$CuO$_4$\,\cite{hill2008observation,ellis2010magnetic}. We will discuss the momentum dependence later in Sec.~\ref{sec:discussion}.

Figure~\ref{fig:rixs} presents the trRIXS spectra for $E_{\rm edge}-30t_h\!\le\!\omega_{\rm in}\!\le\!E_{\rm edge}+10t_h$. This range of incident energy covers the four intermediate-state resonances, labeled by arrows between panels (a) and (b). Let us start with the equilibrium spectrum shown in Fig.~\ref{fig:rixs}(a1). The entire RIXS spectrum is dominated by the Mott-gap excitation at $\omega \gtrapprox 6t_h$ [highlighted to the right of Fig.~\ref{fig:rixs}(a1)]. This excitation has a center at $\omega\!=\!U$ but spreads-out by $\pm 2\,t_h$ due to quantum fluctuation. It reflects the doublon-hole fluctuations across the Mott gap \cite{jia2012uncovering,abbamonte1999resonant,lu2006incident, doring2004shake}. Apart from this high-energy excitation, the equilibrium spectrum displays a low-energy peak at $\omega\sim 1.6t_h$ for the ``poorly-screened'' channel ($\omega_{\rm in}\!\sim\! E_{\rm edge}-10.5t_h$). It corresponds to the bimagnon excitation, where two antiparallel spins flip together and cost energy of $\sim 3J$\,\cite{tohyama2002resonant,nagao2007two,forte2008magnetic,hill2008observation,ellis2010magnetic}. These two main features -- the  Mott-gap and bimagnon excitation -- characterize the energy scales of $U$ and $J$, and reflect the charge and spin excitations we expect to detect in a Mott insulator.

\begin{figure}[!t]
\begin{center}
\includegraphics[width=\columnwidth]{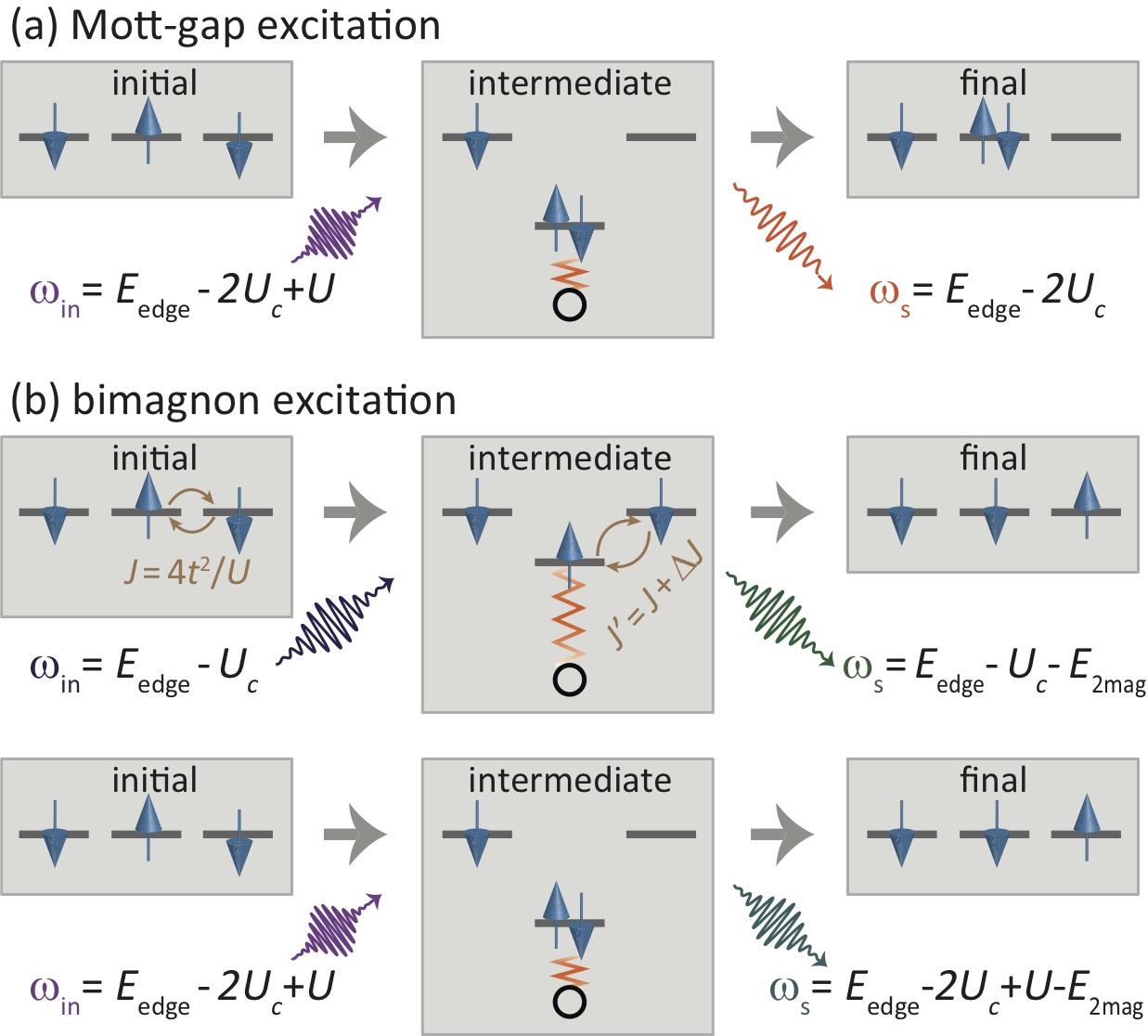}
\caption{\label{fig:RIXSprocesses} Schematic illustrating the two significant excitations revealed by trRIXS: (a) the Mott-gap excitation through the ``well-screened'' initial state and ``double-occupancy'' final state; (b) the bimagnon excitation through the ``poorly-screened'' initial state and ''poorly-screened'' final state, and through the ``well-screened'' initial state and ''well-screened'' final state.
}
\end{center}
\end{figure}

With the accessibility to specific intermediate states, RIXS provides more information about the origin of these excitations in a scattering process. The distribution of equilibrium RIXS intensity roughly follows the \emph{initial-state resonance} profile revealed in the XAS of Fig.~\ref{fig:xasSpec}: The intensity concentrates near $\win=E_{\rm edge}-10.5t_h$ and $E_{\rm edge}-16t_h$ -- the ``poorly-screened'' and ``well-screened'' resonances. Besides, since the second-step de-excitation in RIXS is an x-ray emission process, the RIXS spectrum (of a Mott insulator) also exhibits a slight dependence along the $\wout\! =\! \win\! -\! \omega$ direction, which we refer to as the \emph{final-state resonance}. To better reveal this resonance, we present the trRIXS snapshots also in a ($\wout-\win$) view [see Figs.~\ref{fig:rixs}(b1)]. Affected by the core hole, the final-state resonance is most evident in the ``double-occupancy'' channel (green, $\wout\!\sim\!E_{\rm edge}-22.53t_h$). The combination of these two types of resonance accounts for most spectral weight in RIXS and reveals the origin of various excitations. As explained in Fig.~\ref{fig:RIXSprocesses}, the Mott-gap excitation originates from the combination of a ``well-screened'' (initial-state) x-ray photoexcitation and ``double-occupancy'' (final-state) photo-deexcitation. In contrast, the bimagnon excitation originates from a ``poorly-screened''  photoexcitation and ``poorly-screened'' deexcitation (or a ``well-screened''/``well-screened'' combination), with single occupation in both the initial and final states. The presence of core-hole attraction lowers the energy barrier for a doublon, increasing the effective spin-exchange energy from $J$ to $J^\prime = 4\,t_h^2\,U/(U^2-U_c^2)$ and leaving an excitation induced by neighboring $\mathbf{S}_\ibf\cdot\mathbf{S}_\jbf$ \cite{van2007theory}. Thus, the resonances with intermediate states elucidate the origin of these multi-particle excitations and distinguish them from other excitations at same energy scales, which will be introduced later after a pump.

Out of equilibrium, the spin and charge excitations, including their energy, intensity, and coherence, can be manipulated by the external pump\cite{mentink2015ultrafast, wang2018theory}, which may result in different effective interactions and emergent phases. Note, that the dynamics of the bimagnon excitation at $\qbf\!=\!(0,0)$ have been employed to track the evolution of the transient effective $J$ in time-resolved Raman scattering\,\cite{batignani2015probing,bowlan2018using,wang2018theory}, but the information at finite momentum is accessible only through trRIXS\,\cite{dean2016ultrafast}. Particularly, when the pump field is resonant with the Mott gap, as is the case here, it generates numerous particle-hole excitations \cite{wang2017producing}. As a consequence, the nonequilibrium system suffers from not only a transient renormalization of model parameters, but also the formation of different types of excitations and the redistribution of spectral weight. As we will demonstrate below, trRIXS also gives accessibility to the information of these photoinduced excitations and deciphers their distinct origins.

\begin{figure}[!t]
\begin{center}
\includegraphics[width=\columnwidth]{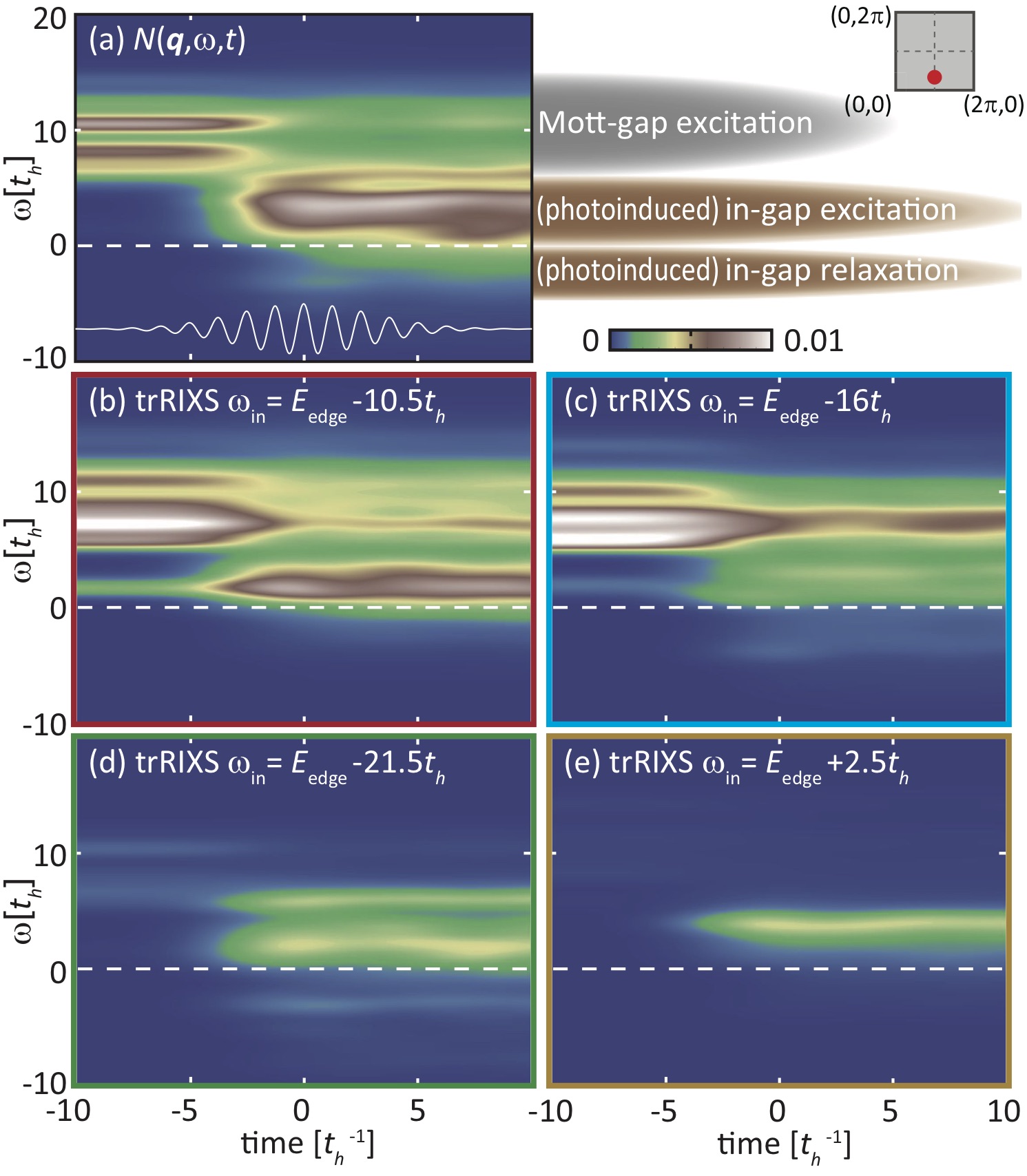}
\caption{\label{fig:RIXSComp} (a) Time evolution of the dynamical charge structure factor $N(\qbf,\omega,t)$ in the same system of Fig.~\ref{fig:rixs} and pump conditions for $\qbf=(\pi,\pi/3)$. The inset denotes the evolution of pump field and the right-hand side markers guide the eye for the Mott-gap excitations and photoinduced excitations. (b-e) The evolution of trRIXS spectrum for four initial-state resonances: (b) ''poorly-screened'', (c) ''well-screened'', (d) ''double-occupancy'', and (d) ''empty-site'' channels.
}
\end{center}
\end{figure}

\begin{figure*}[!t]
\begin{center}
\includegraphics[width=18cm]{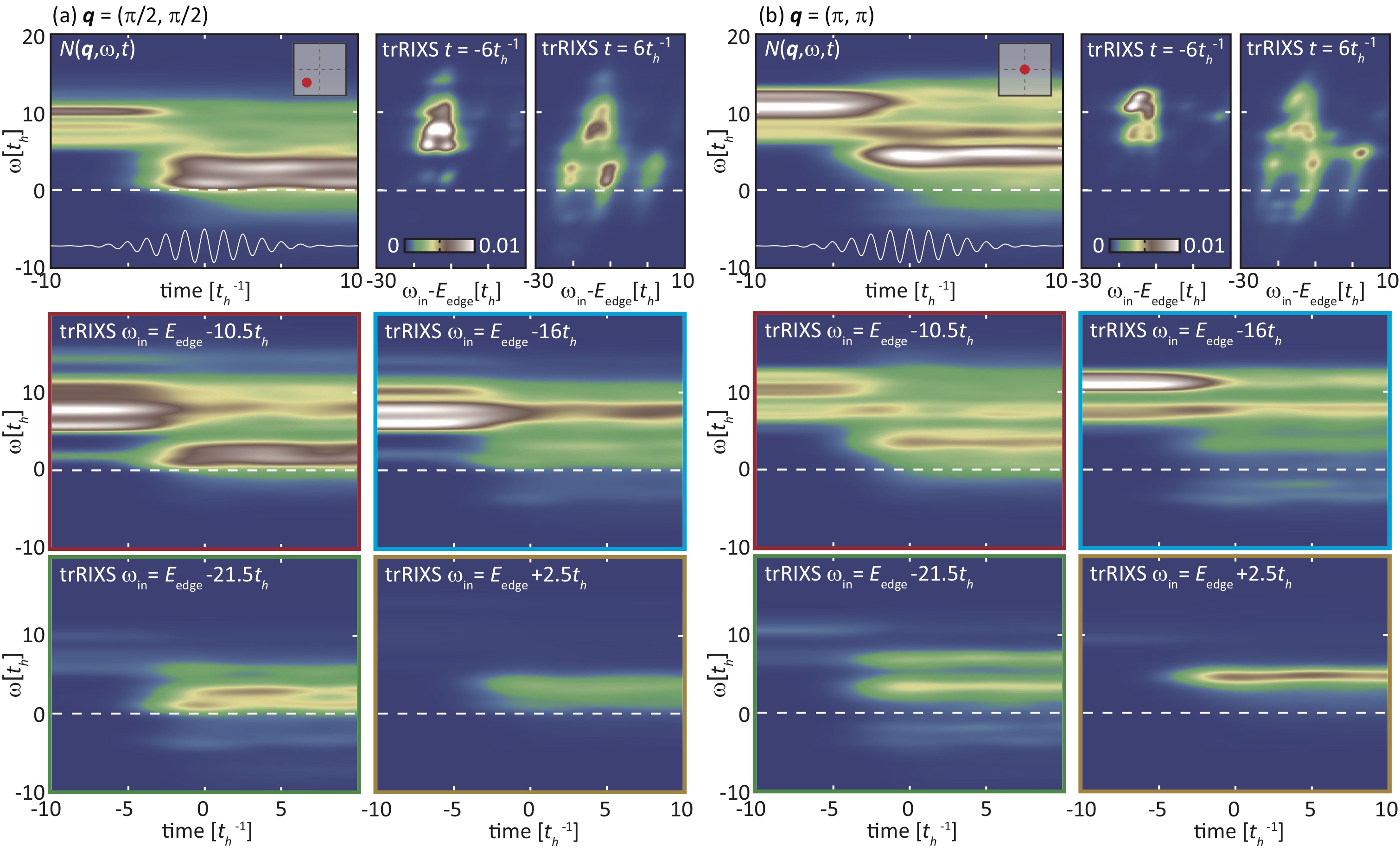}
\caption{\label{fig:RIXSmomDep} Time evolution of the dynamical charge structure factor $N(\qbf,\omega,t)$, trRIXS snapshots before and after the pump, and the time evolution of trRIXS spectra at four resonant $\win$s for (a) $\qbf=(\pi/2,\pi/2)$ and (b) $\qbf=(\pi,\pi)$. The pump condition and the layout are the same as Fig.~\ref{fig:RIXSComp}.
}
\end{center}
\end{figure*}

As shown in Figs.~\ref{fig:rixs}(a3-a6) and (b3-b6), the final-state resonance becomes gradually more evident with the entering of the pump field, exhibiting two ``emission lines''. Notably, the ``well-screened'' final-state resonance (denoted by the blue horizontal arrow) becomes less visible with the growth of the other three resonances. This indicates the unraveling of correlations after a pump: The spin order is reduced compared to the antiferromagnetic (AFM) ground state, resisting a core-hole-trapped doublon from decaying to a singlet by emitting a photon. For the same reason, the magnetic excitation, manifested as the bimagnon here, broadens into a low-energy continuum with the destruction of AFM order [also see Fig.~\ref{fig:RIXSComp}(b)]. This indicates the deconfined motion of a single electron. A similar phenomenon has been observed in the dynamical spin structure factor of a pumped Hubbard model\,\cite{wang2017producing}. At the same time, the Mott-gap excitation gradually loses spectral weight, but keeping the peak position \emph{unchanged}, reflecting the robustness of the Mott gap excitations [also see the dynamical charge structure factor $N(\qbf,\omega,t)$ in Fig.~\ref{fig:RIXSComp}]. Thus, with the quantification through trRIXS spectra, we understand that photodoping a Mott insulator results in a continuous spectral weight transfer from the Mott gap, instead of a gap closure; meanwhile, the magnetic excitations lose coherence, sometimes interpreted as paramagnons\cite{le2011intense, dean2013persistence, dean2013high, lee2014asymmetry, ishii2014high, chaix2018resonant, peng2018dispersion}.

Apart from the change of these two main features, some intensity rises in other regions of the spectrum. Evidently, many in-gap excitations show up for $0\!<\!\omega\!<\!4t_h$. These excitations reside around three (initial-state) resonances ($\win - E_{\rm edge} \sim$ $-21.5t_h$, $-10.5t_h$, and $2.5t_h$) and reflect the motion of the photoinduced carriers. Beyond collection of all in-gap excitations reflected in $N(\qbf,\omega,t)$ [Fig.~\ref{fig:RIXSComp}(a)], trRIXS further dissects these photoinduced in-gap excitations contributed by doublons, holes, and singly-occupied electrons, through the control of the incident energy $\win$. With the pump condition employed in this paper, major in-gap excitations originate from the singly occupied electrons. Thus, the in-gap excitations in the (initial-state) ``poorly-screened'' resonance carry the major spectral weight [see the comparison between Fig.~\ref{fig:RIXSComp}(b) and other panels]. As mentioned above, this deconfined motion of singly-occupied electrons is a consequence of the destruction of AFM order. Other than this channel, the photoinduced doublons (revealed by the ``double-occupancy'' channel) and holes (revealed by the ``empty-site'' channel) also contribute to the in-gap excitations [see Figs.~\ref{fig:RIXSComp}(d) and (e)]. With a negative $t_h^\prime$ in our model Hamiltonian Eq.~\eqref{eq:Hubbard}, the particle-hole symmetry is broken and the motion of doublons involves more correlation with magnons than that of holes.\,\cite{hanke20103,moritz2011investigation, jia2014persistent, wang2014real, li2016signatures, parschke2019numerical} This effect leads to the broad distribution of spectral weight for $\win=-21.5t_h$ and coherent excitation for $\win=2.5t_h$, consistent with the trend in equilibrium electron and hole doping\cite{jia2012uncovering}.

The above evolution of the Mott-gap, bimagnon and various photoinduced in-gap excitations reflects the continuous photodoping away from a half-filled Mott insulator. Since the pump frequency is resonant to the Mott gap, the photodoping effect overwhelms the transient renormalization of model parameters, although a slight softening of the bimagnon peak is still visible. A more evident engineering of the spin-exchange energy and magnetic excitations requires a pump off resonant with the upper Hubbard band\cite{wang2018theory}.

Besides the Stokes responses above the elastic line, the trRIXS spectrum of the post-pump system displays several anti-Stokes features, present also in the $N(\qbf,\omega,t)$. Since both the ``double-occupancy'' and ``well-screened'' channels involve double-occupied intermediate states, they contribute to the major anti-Stokes relaxation for $-4t_h<\omega<0$. Physically, the relaxation reflects the process that the pump-induced excited states release energy, by emitting a photon with $\wout>\win$. Note that these anti-Stokes responses are associated with specific coherent processes and cannot be simply attributed to effective heating, in a strong pump regime far beyond the linear-response\cite{wang2018theory}. The appearance of these new spectral features reflects the nature of the many-body excited states stimulated by the pump field.

Taking all incident and scattering photon energies into consideration, the post-pump trRIXS spectrum roughly presents a checkerboard at the intersections between the initial- and final-state resonances [see Figs.~\ref{fig:rixs}(a6) and (b6)]. The collective excitations with momentum transfer $\qbf$ and specific energy loss $\omega$ can be extracted from the positions and intensities of these spectral features. Therefore, trRIXS enables the above dissection of these excitations as well as their underlying physical processes, through the tunability of the probe frequency $\win$.

\section{Discussions}\label{sec:discussion}
Using a fixed pump-probe condition and a momentum transfer $\qbf=(\pi,\pi/3)$, we have analyzed the spectral features of trRIXS and their origin in the previous section. Compared to time-resolved optical spectroscopy and Raman scattering, the advantage of trRIXS lies in not only the tunability of intermediate states, but also the accessibility of finite momentum. In this section, we will discuss the momentum dependence of trRIXS spectra. To provide guidance for future experiments, we will also discuss the influence of two crucial time scales, the core-hole lifetime $\tau_{\rm core}$ and width of the probe pulse $\sigma_{\rm pr}$.

\subsection{Momentum Dependence of trRIXS Spectra}\label{sec:momDep}
Figure~\ref{fig:RIXSmomDep} shows the $N(\qbf,\omega,t)$ and trRIXS spectra for another two important high-symmetry momenta, $\qbf=(\pi/2,\pi/2)$ and the AFM wavevector $\qbf=(\pi,\pi)$. The calculations for other momenta are presented in the Appendix. The spectral distribution in the ($\wout-\win$) view is consistent with our association of spectral features in Sec.~\ref{sec:trRIXSres}: the Mott-gap and bimagnon [absent for $\qbf=(\pi,\pi)$; see below] excitations dominate the equilibrium spectra; the pump field induces in-gap excitations clustered at the ``poorly-screened'', ``double-occupancy'' and ``empty-site'' resonances; the final-state resonance becomes more evident after the pump. The association of these spectral features and their emergence after pump are independent of the momentum transfer.

The accessibility of various momentum further reveals the dispersion and spectral weight distribution of these collective excitations. As shown in both the $N(\qbf,\omega,t)$ and corresponding trRIXS spectra, the Mott-gap excitation hardens for larger momenta. This energy shift is on the order of 2$t_h$ and is caused by the dispersion of the single-particle band structure on top of the Mott gap. Moreover, the induced photoinduced in-gap excitations are also dispersive: softening near the AFZB where spin excitations (magnons) are costly; hardening at larger momenta where gapless magnons dominate. Though our calculation is based on a Mott insulator, the dispersion of these in-gap charge excitations would also be important for tracking the evolution of charge orders in a CDW or stripe phase\cite{mitrano2019ultrafast, mitrano2019evidence}.

In contrast to both $(\pi,\pi/3)$ and $(\pi/2,\pi/2)$, the bimagnon excitation is less visible far from the AFZB and is absent for $\qbf=(\pi,\pi)$ trRIXS spectrum. That is because the formation of bimagnons requires the anti-alignment of two spins. This phenomenon has been observed previously in equilibrium\cite{hill2008observation,ellis2010magnetic}. Other than the change of spectral weight, the bimagnon is almost dispersionless, consistent with previous experimental results\cite{chaix2018resonant}. Due to its momentum independence, the bimagnon energy provides a more convenient characterization of transient spin exchange energy, compared to gapless and dispersive magnon excitations.

\subsection{Impact of the Core-Hole Lifetime and the Probe Pulse}
\begin{figure}[!t]
\begin{center}
\includegraphics[width=\columnwidth]{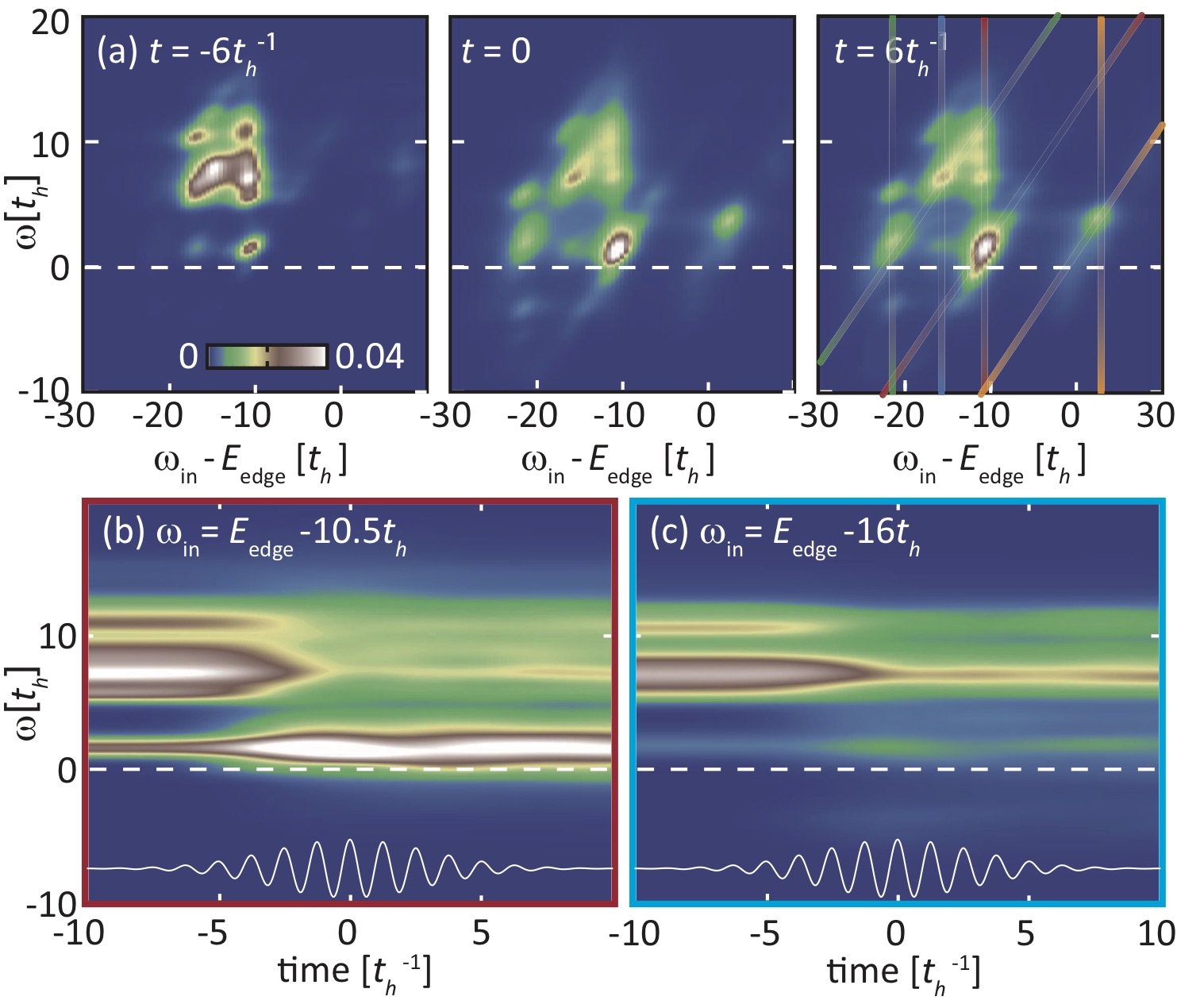}
\caption{\label{fig:RIXSlifetime}  (a) Three snapshots of the $\qbf=(\pi,\pi/3)$ trRIXS spectra with the same pump and probe condition as Fig.~\ref{fig:rixs} but a \emph{longer core-hole lifetime} of $\tau_{\rm core} = t_h^{-1}$. The white dashed lines denote the elastic response, while the colored lines represent the resonance for initial and final states in the same manner as Fig.~\ref{fig:rixs}.
(b) Time evolution of the trRIXS spectra system for two initial-state resonances: (b) ''poorly-screened'' channel and (c) ''well-screened'' channel.
}
\end{center}
\end{figure}

There are a few important time scales that determine the properties of a pump-probe spectroscopy. Other than the intrinsic valence electron time scales set by model parameters $t_h^{-1}$ and $U^{-1}$, which determine the elementary excitations to be detected, the core-hole lifetime and the probe width are also crucial for a trRIXS spectrum. A proper consideration of these timescales gives an expectation of experimental data quality. 

In terms of the perturbation order, the excitations revealed by trRIXS can be classified into linear and nonlinear. The linear excitations include the Mott-gap excitation, photoinduced in-gap excitations, and anti-Stokes relaxations, which are accessible in $N(\qbf,\omega,t)$. Through the resonant intermediate state, trRIXS spectra access nonlinear excitations, including the bimagnon and high-order corrections, which vanish in $N(\qbf,\omega,t)$. As a photon-in-photon-out scattering process, these nonlinear effects are consequences of the finite lifetime $\tau_{\rm core}$ of the intermediate state: In the ultrashort core-hole lifetime limit ($\tau_{\rm core}t_h\ll 1$), the trRIXS spectrum faithfully depicts the $N(\qbf,\omega,t)$; with the increase of $\tau_{\rm core}$, the nonlinear effects including the bimagnon excitations are expected to become more evident\,\cite{tohyama2018spectral}. To demonstrate this lifetime dependence, we study the trRIXS spectra with a lifetime $\tau_{\rm core} = t_h^{-1}$, twice of that in the above results [see Fig.~\ref{fig:RIXSlifetime}]. In addition to the increase of intensity with roughly the square of the ratio between lifetimes, the nonlinear part of the spectrum, the bimagnon excitation, becomes more evident. At the same time, the linear part, the Mott gap excitation, still faithfully reproduces the $N(\qbf,\omega,t)$ [see the comparison with Fig.~\ref{fig:RIXSComp}. Interestingly, due to a greater contribution of the intermediate state, the resonant effects of both $\win$ and $\wout$ are clearer. We want to emphasize that though the core-hole lifetime acts as a shape factor in Eq.~\eqref{eq:RIXScrossSec3}, it reflects the intrinsic properties of the material and cannot be controlled by the external field.

\begin{figure}[!t]
\begin{center}
\includegraphics[width=\columnwidth]{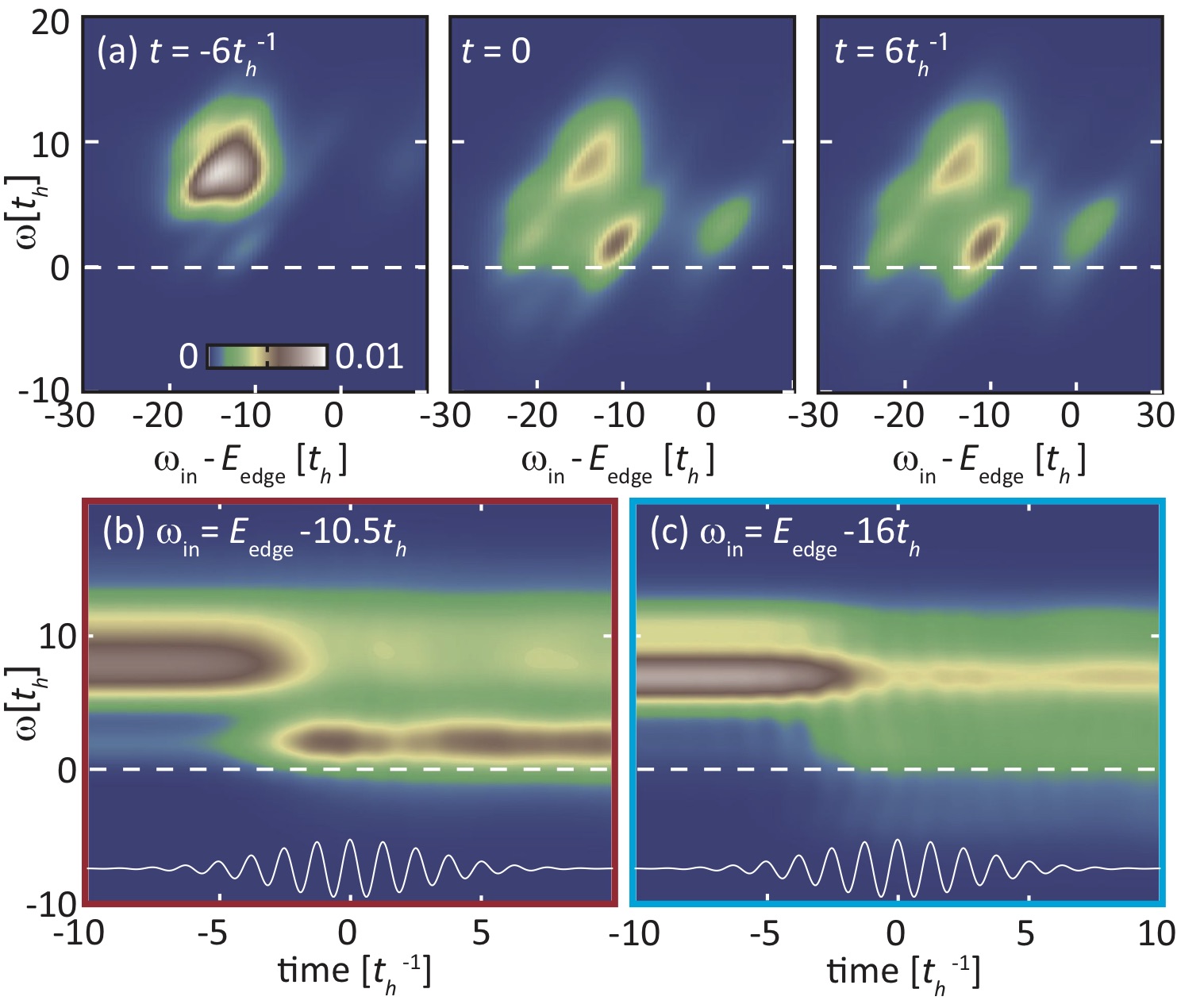}
\caption{\label{fig:RIXSprobeWidth}  (a) Three snapshots of the $\qbf=(\pi,\pi/3)$ trRIXS spectra with the same pump condition as Fig.~\ref{fig:rixs} but a \emph{shorter probe width} of $\sigma_{\rm pr} = 0.5t_h^{-1}$. The white dashed lines denote the elastic response.
(b) Time evolution of the trRIXS spectra system for two initial-state resonances: (b) ''poorly-screened'' channel and (c) ''well-screened'' channel.
}
\end{center}
\end{figure}

\begin{figure*}[!t]
\begin{center}
\includegraphics[width=16cm]{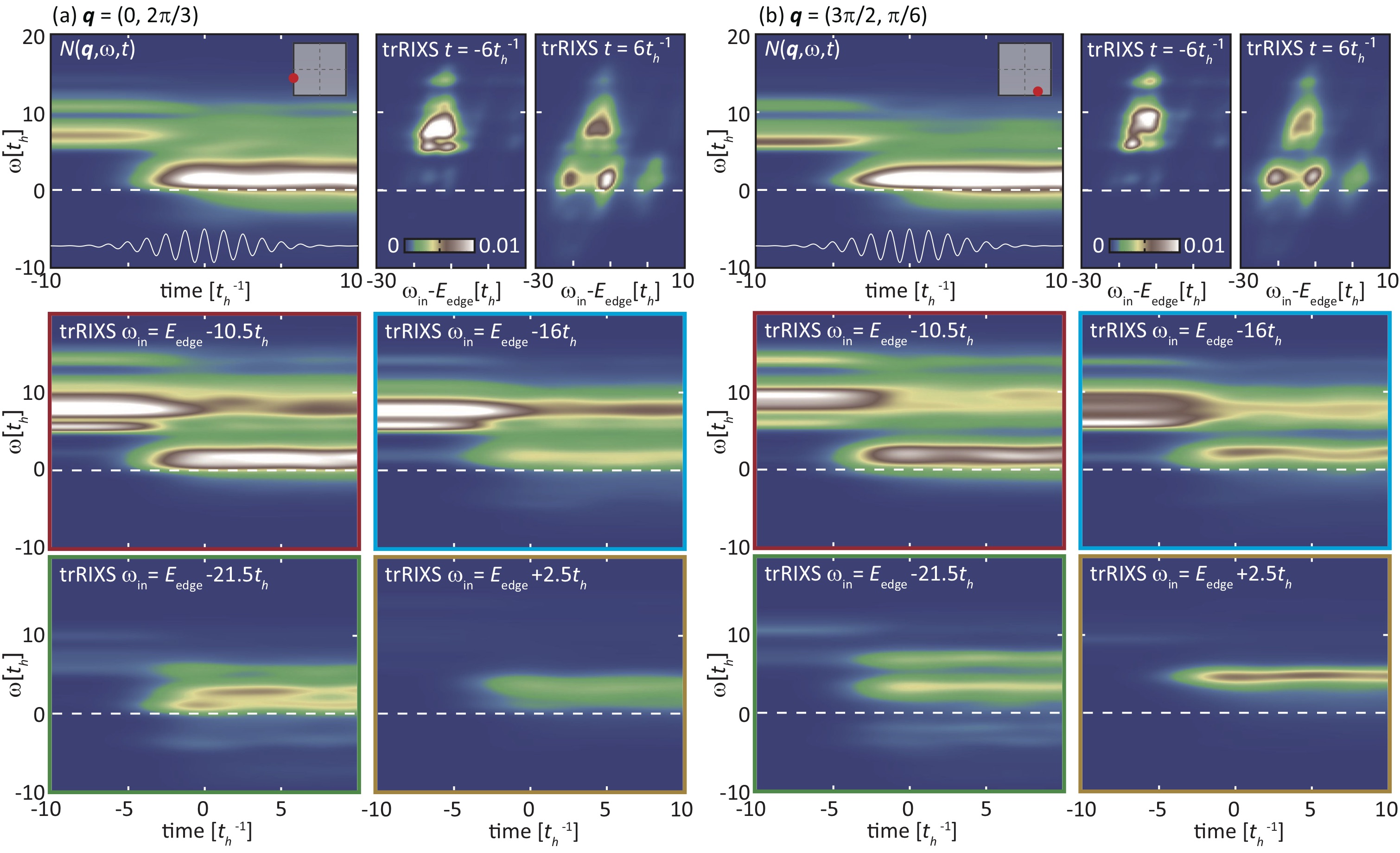}
\caption{\label{fig:RIXSOther1} Time evolution of the dynamical charge structure factor $N(\qbf,\omega,t)$, trRIXS snapshots before and after the pump, and the time evolution of trRIXS spectra at four resonant $\win$s for (a) $\qbf=(0,2\pi/3)$ and (b) $\qbf=(\pi/2,\pi/6)$. The pump condition and the layout are the same as Fig.~\ref{fig:RIXSComp}.
}
\end{center}
\end{figure*}

The appearance of nonlinearity also depends on the third time scale in a pump-probe experiment -- the width of the probe pulse $\sigma_{\rm pr}$. In contrast to the intrinsic $\tau_{\rm core}$, the $\sigma_{\rm pr}$ can be controlled by the light source in an experiment, which balances the energy resolution and time resolution. Such a balance should be considered specifically for different systems. For the strongly correlated quantum materials such as cuprates, due to the bimagnon energy at $\sim t_h$, the ideal probe width to distinguish this feature from the elastic peak (excluded in the calculation but should be present in experiments) should be larger than $t_h^{-1}\sim 15$\,fs.  To investigate the impact of the probe width, we present the trRIXS results of $\sigma_{\rm pr}=0.5t_h^{-1}$ in Fig.~\ref{fig:RIXSprobeWidth}, in contrast to the $1.5t_h^{-1}$ adopted in Fig.~\ref{fig:rixs}. As we expected, the bimagnon peak is less evident for such a narrow probe. Interestingly, since $\sigma_{\rm pr}$ is even smaller than the period of the pump field $2\pi/\Omega = 1.26t_h^{-1}$, the trRIXS spectra capture the fluctuations induced by the periodic gauge field.\cite{chen2019theory} This extreme should be observable in terahertz-pump experiments, which becomes feasible using the LiNbO$_3$, organic crystals, and synchrotron generators in recent years\cite{dienst2011bi,wu2013intense,ruchert2013spatiotemporal}. Besides, the recent development of ultrashort sub-femtosecond x-ray pulses\cite{huang2017generating} enables the observation of fast quantum fluctuation in electronic states with well-separated energy scales (e.g.~small molecules).

\begin{figure*}[!ht]
\begin{center}
\includegraphics[width=16cm]{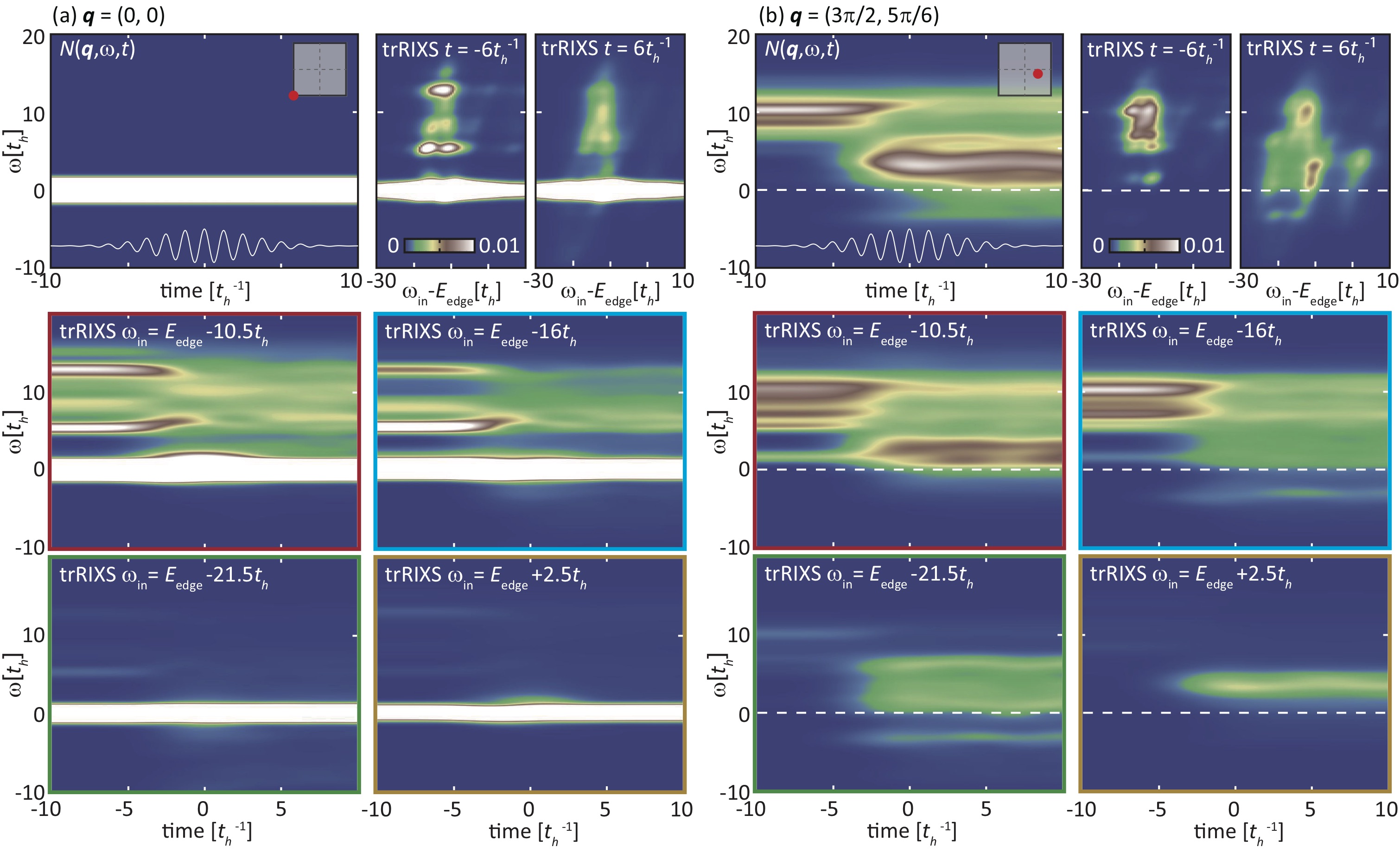}
\caption{\label{fig:RIXSOther2} Time evolution of the dynamical charge structure factor $N(\qbf,\omega,t)$, trRIXS snapshots before and after the pump, and the time evolution of trRIXS spectra at four resonant $\win$s for (a) $\qbf=(0,0)$ and (b) $\qbf=(\pi/2,6\pi/6)$. The pump condition and the layout are the same as Fig.~\ref{fig:RIXSComp}.
}
\end{center}
\end{figure*}

\section{Summary and Outlook}\label{sec:summary}
To summarize, we have reported an unbiased trXAS and trRIXS study in a correlated system using time-dependent exact diagonalization. With spectral weight transfer among well-separated peaks, the trXAS provides a convenient visualization of photodoping. On top of it, the trRIXS spectrum reflects the evolution of charge and spin excitations, including their energy, intensity, dispersion, and coherence. In  a Mott insulator, the combination of both spectra characterizes the evolution of Mott-gap and bimagnon excitations. The accessibility to these transient excitations and their momentum dependence provides an indispensable tool to uncover the gradual transition while doping a Mott insulator and the collective excitations associated with ultrafast emergent phenomena.

Furthermore, the tunability of the incident photon frequency near the x-ray edge provides element selectivity for trRIXS, through which one can dissect collective excitations associated with different intermediate electronic states. From this aspect, trRIXS partially reveals the single-particle information, which is crucial for materials where trARPES is inaccessible, such as under high pressure or on dirty surfaces.

In this paper, we have considered the simplest model for correlated systems and conditions for trRIXS, due to the numerical complexity of the four-time correlation functions. With more computational effort, the same calculation can be extended into direct trRIXS (such as Cu $L$-edge) to detect the evolution of magnon/paramagnon excitations. Besides, the long-time relaxation back to equilibrium is not considered in our calculation using a microcanonical ensemble. Therefore, we restrict our analysis to ultrafast dynamics during, or shortly after, the pump. To mimic the dissipation, a canonical approach is solving the master equation with a Lindblad dissipator. This also requires the spectra of Eqs.~\eqref{eq:XASexpr} and \eqref{eq:RIXScrossSec3} be rewritten as ensemble expectation values evaluated by the density matrix.

\section*{Acknowledgement}
We thank Y. Peng for insightful discussions and F. Liu for technical support.
Y.W. acknowledges the Postdoctoral Fellowship in Quantum Science of the Harvard-MPQ Center for Quantum Optics. Y.C., C.J.J, B.M., and T.P.D acknowledge support from the U.S. Department of Energy, Office of Science, Office of Basic Energy Sciences, Division of Materials Sciences and Engineering, under Contract No.~DE-AC02-76SF00515. This research used resources of the National Energy Research Scientific Computing Center (NERSC), a U.S.~Department of Energy Office of Science User Facility operated under Contract No.~DE-AC02-05CH11231.

\appendix
\section{Time-Resolved RIXS Spectra for Other Momenta}

In Secs.~\ref{sec:trRIXSres} and \ref{sec:momDep} of the main text, we have discussed the momentum dependence of trRIXS spectra with a focus on the $\qbf =(\pi,2\pi/3)$, $(\pi/2,\pi/2)$, and $(\pi,\pi)$. Here, we present the spectra for the other four independent momenta in the 12D Betts cluster. 

Figure \ref{fig:RIXSOther1} shows two momenta near the antinode $\qbf=(\pi,0)$. Their spectral distribution and pump-induced dynamics are similar to that of $\qbf=(\pi,\pi/3)$ [see Fig.~\ref{fig:RIXSComp}. This similarity is reflected in the $N(\qbf,\omega,t)$ and the Mott-gap excitations. Notably, the bimagnon features in these two momenta are much weaker compared to the $\qbf=(\pi,\pi/3)$ result in the main text, because these two momenta reside outside the AFM zone, where spin correlations rapidly drop.
 
The other two momenta, the $\qbf=(0,0)$ and $(3\pi/2,5\pi/6)$ are shown in Fig.~\ref{fig:RIXSOther2}. Evaluated from the cross-section of Eq.~\eqref{eq:RIXScrossSec3}, the RIXS spectrum for $\qbf=(0,0)$ is dominated by an elastic peak. This elastic response reflects the total charge of the valence band. Due to the commutation of $\rho_\qbf$ and the Hamiltonian Eq.~\eqref{eq:Hubbard}, the dynamical charge structure factor $N(\qbf,\omega,t)$ displays nothing other than this elastic peak. However, trRIXS reflects richer information, including the Mott gap at small momenta. The bimagnon excitation does not vanish but is overwhelmed by the strong elastic peak. Experimental extraction of this low-energy mode requires a second-derivative analysis.

\bibliography{paper}

\begin{thebibliography}{89}
\bibitem[{flu()}]{fluence}
\bibinfo{note}{The fluence is determined by $I=\sqrt{\pi} h c \sigma \epsilon_0
  A_0^2\Omega^2 /2 e^2a_0^2$, where $e$ is the elementary charge, $\epsilon_0$
  is the vacuum dielectric constant, $h$ is the Planck constant, $c$ is the
  speed of light, and $a_0$ is the lattice constant (taken as $3.78$\AA\ here).
  The evaluation of fluence has taken the Gaussian envelope of pump pulse in to
  consideration.}
\end{thebibliography}
\end{document}